FAST POINT-FEATURE LABEL PLACEMENT

FOR DYNAMIC VISUALIZATIONS

By

Kevin Dean Mote

A thesis submitted in partial fulfillment of
the requirements for the degree of

MASTER OF SCIENCE IN COMPUTER SCIENCE

WASHINGTON STATE UNIVERSITY
Department of Computer Science

DECEMBER 2007

To the Faculty of Washington State University:

The members of the Committee appointed to examine the thesis of Kevin Dean Mote find it satisfactory and recommend that it be accepted.

_______________________________________

Chair

_______________________________________

_______________________________________

FAST POINT-FEATURE LABEL PLACEMENT

FOR DYNAMIC VISUALIZATIONS

Abstract

Kevin Dean Mote, M.S.
Washington State University
December, 2007

Chair: John H. Miller

This paper describes a fast approach to automatic point label de-confliction on interactive maps. The general Map Labeling problem is NP-hard and has been the subject of much study for decades. Computerized maps have introduced interactive zooming and panning which has intensified the problem. Providing dynamic labels for such maps typically requires a time-consuming pre-processing phase. In the realm of visual analytics, however, the labeling of interactive maps is further complicated by the use of massive datasets laid out in arbitrary configurations, thus rendering reliance on a pre-processing phase untenable. This paper offers a method for labeling point-features on dynamic maps in real time without pre-processing. The algorithm presented is efficient, scalable, and exceptionally fast; it can label interactive charts and diagrams at speeds of multiple frames per second on maps with tens of thousands of nodes. To accomplish this, the algorithm employs a novel geometric de-confliction approach, the "trellis strategy," along with a unique label candidate cost analysis to determine the "least expensive" label configuration. The speed and scalability of this approach make it well-suited for visual analytic applications.

# Table of Contents

# TABLE OF FIGURES

# LIST OF TABLES

# 1. INTRODUCTION

The problem of placing non-conflicting labels on maps is simple in its specification: attach a label unambiguously to every point or feature on a map without allowing the labels to overlap each other or other features. Although it is theoretically feasible to solve the problem by simply enumerating all possibilities, the combinatorial explosion of this approach renders it intractable. Many different variants of the label placement problem have in fact been shown to be NP-hard [2, 15, 27], even in the one-dimensional case [35]. Research is therefore directed at finding the best approximation of an optimal solution. A decade ago the problem was identified as one of the most important areas of research in Discrete Computational Geometry [4]. It has long been a critical issue for such endeavors as graph and map drawing, as well as geographic and avionic information systems.

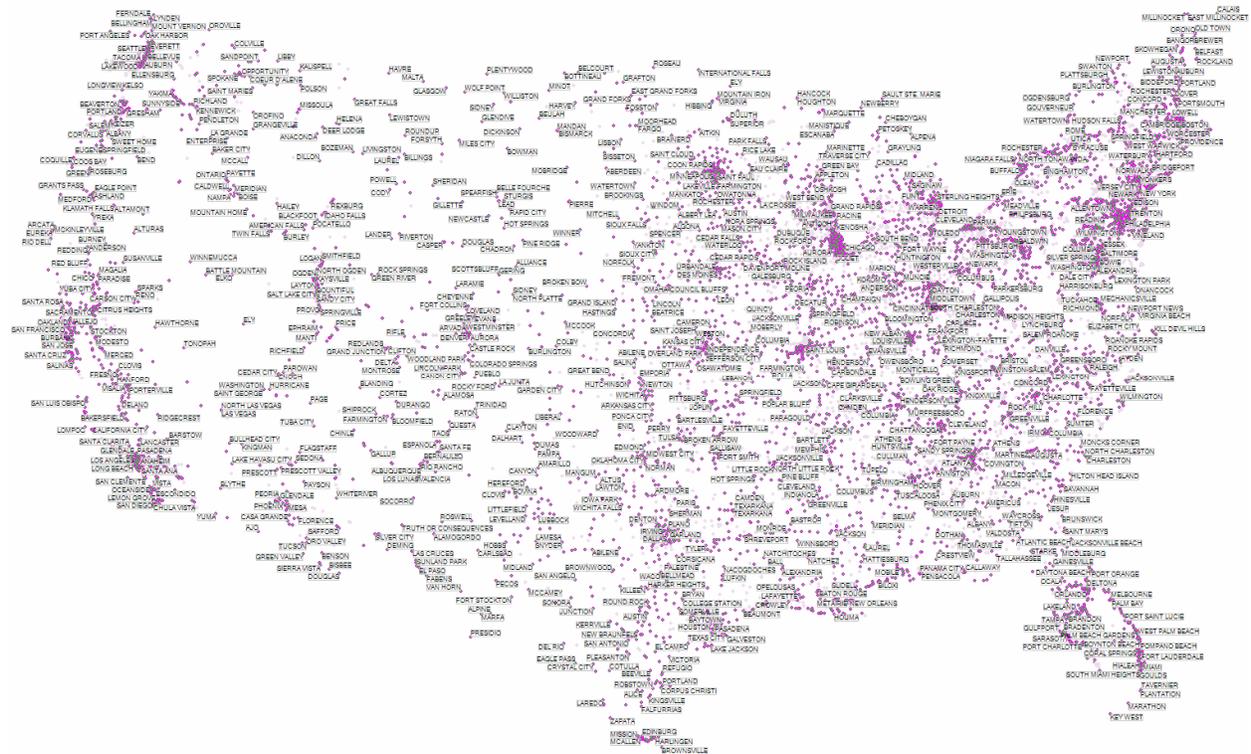

**Figure 1: A view of over 10,000 cities in the U.S. displaying 800 labels that were computed in less than 0.07 seconds.**

With the advent of information visualization and visual analytics, the challenge has only intensified. Visual analytic software brings unique demands: agile interactivity (including zooming and panning in 2D, or even rotation in 3D), unpredictable configurations (produced on the fly), and immense feature sets (multiple thousands of nodes). By way of example, the Starlight Visual Information System can generate a variety of models, or "views", consisting of multiple feature points displayed graphically in clusters, network graphs, or geospatial distributions, etc.[34] These views, generated by user-specified queries in a highly interactive environment, can easily contain tens of thousands of points or more. Besides Starlight, many other systems exist which spatially represent the text documents of large databases (e.g.,[37, 40]). In such environments the need to preserve contextual awareness through rapidly produced and readable labels is critical. But despite many recent advancements, most computational solutions for automatic label placement have proved inadequate for the exacting demands of interactive and dynamic visualizations.

The general problem of Map Labeling can be sub-divided into various categories, depending on the type of feature being labeled, whether points [5, 36], lines [22, 45], or polygonal areas [11, 23]. This paper will address the point label problem for dense point-clouds (cf. Figures 1 & Figure 17). The remainder of the paper is organized as follows: Section 2 provides a detailed overview of the literature. Section 3 specifies the precise problem. Section 4 describes the algorithmic solution of this paper. Section 5 outlines the results obtained by applying the algorithm to maps with dense feature sets. Section 6 summarizes the findings and offers suggestions for further study. Finally, a thorough description of the author's implementation is described in Appendix A.

## 2. HISTORY OF PROBLEM AND RELATED WORK

Research on the problem of automatic label placement stretches back for decades and has proceeded along a number of parallel tracks; researchers in cartography, computational geometry, and geographic information systems have been the principal developers. Accordingly, a wide variety of strategies have been applied over the years: from greedy and exhaustive rules-based approaches, to "divide and conquer," gradient descent, simulated annealing, genetic algorithms, linear/integer programming techniques, tabu search, ant colonies, and many more. In addition, the research has addressed various different label types in regard to their size, shape, and configuration. For example, labels have been modeled as squares, circles, fixed-height rectangles, or elastic frames. Other parameters to the problem include label orientation (axis-aligned or arbitrary), the number of label candidates per feature (2, 4, 8, or more), candidate space configuration (fixed, slider, or non-adjacent with leader-lines), optimization goal (size- or count-maximization), metric of success (speed or thoroughness), user-dependence (automated or guided), and application-domain (cartography, geographic information systems, information visualization systems). The influence of cartographic principles on information visualization is discussed in [37]. For comprehensive information on map labeling, along with references to the approaches mentioned above, see Wolff's excellent website and associated bibliography [44].

### 2.1. BRIEF OVERVIEW

Historically, most of the literature has centered on the cartographic problem of finding the *best* solution: producing a map with the most labels, for example. Little consideration was given to the *speed* of the algorithm. Within the past few years, however, new breakthroughs have been made to provide faster algorithms for label generation. This has become increasingly important in the age of dynamic, computer-generated maps and displays. For example, until a few years ago, the best solutions in the scientific literature operated on the order of seconds or minutes, which is acceptable when generating a static map for printing. It is too slow, however, for dynamic maps that allow zooming and panning. In these applications, the label positions must be re-calculated with every change of scale or scope. An acceptable algorithm, therefore, must operate in real-time (i.e., "on the wheel of a mouse") for such an application to be useful.

Wagner *et al*. were among the first to provide a faster approach, suggesting heuristic rules that brought a significant improvement over previous approaches but that were still too slow for large sets, measuring in minutes for large sets [43]. Petzold followed by providing real-time labeling, albeit with a high pre-processing cost [30]. A so-called "Real-Time Method," intended to be suitable for personal navigation applications, was described in [48], but the authors concede that "for real-time applications . . . we find our current implementation not efficient enough." In a related paper an algorithm is presented that is indeed "fast enough for  most real-time map applications," however the authors offer this efficiency only for datasets with about 100 points[19]. A recent approach utilizing a "greedy randomized adaptive procedure" provides good results but requires well over a minute for datasets of 1000 points (on a Pentium IV, 2.8 GHz) and does not test sets larger than this [8]. Most promising, perhaps, is a graph-theoretic algorithm described in [36] that requires $O(n\sqrt{n})$ time, worst case. The authors provide tabulated results indicating a sub-second speed for instances of size 1000. Much larger sets require multi-second time, however, and the approach does not appear to lend itself to feature prioritizing or label placement preference.

The following sections will provide a more detailed look at these and other approaches. Much of this work may be categorized in terms of the optimization strategy employed. It has been common since the earliest days of computational map labeling to approach the problem as an exercise in combinatorial optimization. As such, an objective function is sought whereby the specific cost of a given algorithm may be evalutated. In this respect there are two fundamental approaches by which most algorithms may be categorized. One approach is to view it as a *minimization* problem, whose goal is to minimize the *cost* of the objective function. The second approach is to view it as a *maximization* problem. In this case the algorithm would seek to maximize (a) the *count* of fixed-size labels or (b) the *size* of a fixed count of labels in that labeling. This latter approach is equivalent to finding the smallest factor by which the map has to be "zoomed" such that each point has an unobscured label assigned to it. This was originally proposed by Forman *et al.*[15] who were motivated by finding the maximum font size for a given map. They found an $O(n \log n)$ time algorithm that guarantees a square label size at least half the optimum. They also showed that no better approximation is possible unless P=NP. Similarly, Doddi, *et al.*[9] also approximate the size of labels, but generate their solution in terms of label shapes, allowing circles, non-oriented rectangles, ellipses, etc.

### 2.2. MINIMIZATION STRATEGIES

Two of the earliest algorithms illustrate the cost minimization approach. Christensen, *et al.* [5] describe a so-called "Discrete Gradient Descent" (DGD) approach. In this strategy, all labels are placed in random positions. Each label is then re-located iteratively to one of its alternate positions and the objective function is re-calculated. The repositioning that offers the most improvement is then selected. The major weakness of the DGD algorithm is its inability to escape from local minima of the objective function. In other words, configurations are often encountered wherein a conflict between two labels cannot be resolved without making two or more consecutive steps, at least one of which *increases* the cost of the objective function.

Motivated at least in part by the shortcomings of DGD, a slightly more sophisticated alternative arose, known as "simulated annealing" (SA). Advanced in 1983 by Kirkpatrick, *et al.* [25], simulated annealing is a generic probablistic algorithm which derives its name and inspiration from the process of annealing in metallurgy. This material technique seeks to reduce the defects in a metallic or glass substance by repeatedly heating and cooling the material. The heat excites the atoms of the substance, thereby dislodging them from their original location. A slow cooling then allows them to settle into a (presumably) lower-energy configuration. This process can be compared to repeatedly jostling a half-empty box of sugar cubes in an attempt to coerce them into an arrangement of ordered rows at the bottom of the box. In an analogous way, the SA algorithm begins with a random configuration of labels (the "high-energy" or "high-temperature" configuration). Using the same process as the DGD algorithm, the labeling is allowed to "settle." Unlike that previous strategy, however, local minima are escaped by a stochastically determined allowance of movement in directions other than that of the gradient. In other words, the solution is periodically allowed to get worse rather than better. The amount by which the algorithmic results are allowed to degrade is controlled by a parameter $T$, called the temperature, which decreases over time according to an annealing schedule. Indeed, it can be shown that, for any given finite problem, the probability that the SA algorithm will terminate with the optimum solution approaches 1 as the annealing schedule is extended. This theoretical result is not particularly helpful, however, because the annealing time required to ensure a significant probability of success will usually exceed the time required for a complete search of the solution space. Edmonsen *et al.* [13] (cf. [23]) describe one of the first efforts to use SA to find solutions for the general labeling problem (i.e. point, line, and area features), separating the cartographic knowledge needed to recognize the best label positions from the optimizaton procedure needed to find them.

A decade ago, in an extended set of experiments designed to compare what were then the leading approaches (including all those discussed so far in this paper), Christiansen, *et al.*[6] determined that the SA algorithm outperformed nearly all other approaches both in speed and quality. They noted that for time-critical applications the annealing schedule can be shortend or eliminated altogether while still providing reasonable solutions. They noted that SA had the added advantage of being one of the easiest algorithms to implement.

Having said that, however, it soon became apparent among the researchers that for real-time applications, SA and DGD algorithms simply were not fast enough to be adequate.

### 2.3. MAXIMIZATION STRATEGIES

Moving away from the cost minimization strategy, Agarwal, *et al.*[1] approached the task by observing that the label problem can be formulated equivalently to the *maximum independent set* problem. (In graph theory an independent set is a set of non-adjacent vertices, having no edges connecting any pair of them. The largest such set is the "maximum".) The authors described a simple divide and conquer algorithm for computing a large independent set in a set $R$ on $n$ orthogonal rectangles in the plane. They present a line-stabbing approach that offers a (log $n$)-approximation in $O(n \log n)$ time for rectangles. They define an $\varepsilon$-approximation algorithm as one which returns an independent set of size at least $\gamma/\varepsilon \mid (\varepsilon > 1)$, where $\gamma$ is the size of the *maximum* independent set of $R$. In addition, in the case that all rectangles in $R$ have the same height (a case which models the constant font-size labeling problem) they describe a $(1 + 1/k)$-approximation with running time of $O(n \log n + n^{2k-1})$ for any $k \geq 1$ (where $k$ represents the number of consecutive lines used in the stabbing routine).

Kakoulis and Tollis[23, 24] offer a more general approach to labeling. One of the innovations presented in their seminal work is that of recasting the labeling problem in terms of graph theory. They introduce the concept of a conflict graph, which models the collisions between conflicting label candidates as a graph whose nodes correspond to labels, and whose edges correspond to intersections between labels. After computing this graph and its connected components they use a somewhat greedy algorithm for independent set maximization to split the components into cliques. They then construct a second graph, a bipartite "matching graph" whose nodes are the cliques of the previous step and the features in the map itself. For every clique that contains a candidate of a given feature there is a corresponding edge in the matching graph joining the clique and the feature. The final labeling is produced through a maximum-cardinality matching. Although the authors do not present any time bounds, it has been noted [43] that depending on how the clique check and the matching algorithm is implemented, their algorithm takes $O(k\sqrt{n})$ or even $O(kn)$ time, where $k$ refers to the number of conflicts.

### 2.4. APPROACHING INTERACTIVE SPEEDS

In subsequent years several more studies continued to approach the problem in similar ways, refining the time bounds or re-casting the problem to address different conditons (such as label sizes or configurations, etc.).[10, 14, 38, 41, 42]  The computational geometry theory was continuing to advance and theoretical runtime was continuing to be trimmed incrementally, providing map-makers with ever-increasing efficiency for their historically time-consuming task. Nevertheless, it was becoming increasingly clear to practitioners within the cartographic and burgeoning GIS communities that the hope of obtaining interactive or real-time applications was still far off. Such speeds would require fundamentally different and novel approaches.

### 2.4.1. WAGNER'S "THREE RULES"

The first significant advance on this front came in the form of a heuristic approach which would separate the geometric part of the problem from the combinatorial part. In a seminal paper entitled "Three Rules Suffice for Good Label Placement"[43] (henceforth referred to as "Three Rules") Wagner *et al.* attack the problem in its full generality and present a set of rules that simplify its combinatorial complexity (represented by a conflict graph of candidates) without reducing the size of an optimal solution. Their algorithm significantly out-performed its predecessors, obtaining results that were similar to simulated annealing but required much less time.

The "Three Rules" algorithm is divided into two phases. (Actually an implied third phase of conflict graph generation is assumed, but not discussed.) In the first phase, a set of rules is applied to all the features, labeling as many as possible and reducing the number of candidates of the others. The authors describe these rules with specificity, but they can be summarized as follows:

1.  If a feature has a non-conflicted candidate, choose it, and eliminate the remaining candidates for that feature.

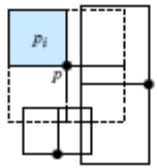

**Figure 2: Wagner's non-conflicted candidates**

2.  If two features are only in conflict with each other, so that all their candidates conflict each other, there will exist two candidates that are mutually independent. Select these two and eliminate the remaining candidates.

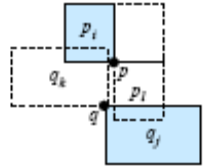

**Figure 3: Wagner's "bonded pairs"**

3. If a feature has only one candidate left, and it overlaps a clique (i.e., a cluster of mutually conflicted candidates), select that candidate and eliminate the others.

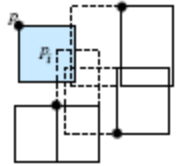

**Figure 4: Wagner's "cliques"**

These rules are applied exhaustively in a recursive manner within the neighborhood of any given feature. The authors are able to show that such an application does not reduce the size of an optimal labeling (as defined by the maximum number of features that can be labeled simultaneously).

The authors then describe a second phase of the algorithm. This phase is applied only when necessary and, while it does not preserve the guarantee of optimality that the first phase promised, it is conceptually simple and practically efficient. In their words: "the intuition is to start eliminating troublemakers where we have the greatest choice." That is to say, they delete the candidates with the maximum number of conflicts for a given feature. They then attempt to re-apply the rules in the neighborhood of the feature. This process is repeated until a solution is found.

Finally the authors demonstrate through direct timed comparison that their approach out-performs their predecessors by a significant amount. In practical terms, the authors have provided what can possibly be considered the first algorithm fast enough to be used by interactive applications such as online-map utilities. This was certainly a remarkable acheivement.

### 2.4.2. PETZOLD'S "REACTIVE GRAPH"

While the "Three Rules" offered a significant advance over all preceding efforts, they still did not provide instantaneous results on large data sets. Yet the GIS community was increasingly interested in finding a solution that could be optimally responsive for the screen-maps that had begun to supercede the traditional paper maps, particularly with respect to mobile applications. Into this arena, Petzold *et al.*[29, 30, 31] offered what they referred to as "stroke-of-a-key" responsiveness. This ground-breaking approach was the first to provide fully-functional zooming and panning in an online environment with (seemingly) instantaneous label deconflictions.

Petzold's profound insight that yielded this breakthrough performance consisted in isolating a sophisticated (and expensive) conflict identification phase and off-loading this burdensome task to a preprocessing phase. This preprocessing step would produce a so-called "reactive conflict graph" augmented by a multidimensional index. This new data structure could then be accessed and exploited during user interaction in an efficient and optimal way.

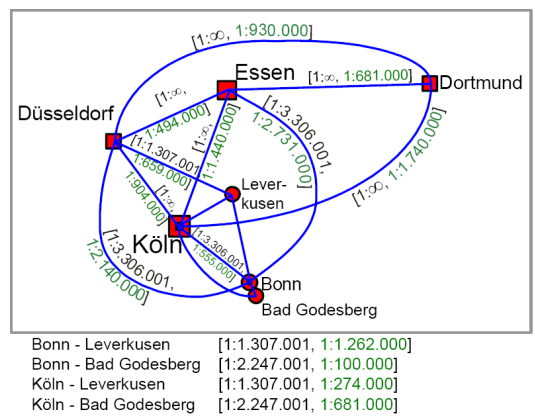

**Figure 5: A model of Petzold's "reactive conflict graph" with attributed scale intervals.**

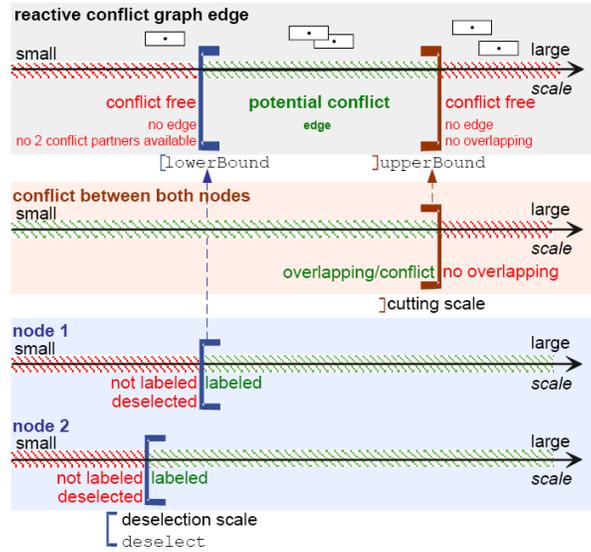

**Figure 6: Modeling an edge in Petzold's "reactive conflict graph". Progressing from "small scale" (on the left) to "large scale" (on the right) represents the action of "zooming in" on an interactive screen map.**

**(1) *Preprocessing Phase.*** As mentioned above, Petzold's algorithm relies on an intensive preprocessing stage to perform the vast majority of the computations necessary for conflict resolution. The goal of this stage is to produce an attributed "reactive conflict graph" as visualized in Figure 5. In this graph, the nodes represent the objects to be labeled and the edges represent potential conflicts between the nodes. A pair of nodes has a "potential conflict" when their corresponding label spaces overlap at any scale. The edges are then attributed with a scale interval describing the precise occurrence of the potential conflict, in terms of map scales. (Note: the "scale" of a map grows *larger* as the user zooms in). This interval has two bounds, as illustrated in Figure 6. The upper bound of the interval (the right bound in Figure 6) is the largest scale where a conflict occurs. Petzold refers to this as the "cutting scale." The idea here is that once a user has zoomed in to such a scale, the distance between the two map objects has grown sufficiently to label both of them; hence, there is no conflict at this scale. The lower bound of the interval (on the left in Figure 6) is the smallest scale where a conflict occurs. Petzold calls this the "deselection scale." This bound accounts for the fact that at certain small scales, some objects are simply not labeled at all. The scale at which a given object is "deselected" is determined as a function of the relative priority of that object. (More specifically, Petzold refers to this function as the "labeling difficulty" of a given object.) Below this lower bound the conflict dissolves since one of the conflict partners is no longer labeled.

A thorough evaluation of the entire set of features (and the edges between every pair) is, of course, an expensive task. Although Petzold offers some algorithmic short-cuts to abbreviate this process, the preprocessing phase of his approach is an intensive undertaking, requiring multiple minutes. However, the termination of this process yields a populated data structure that can be exploited efficiently in the subsequent phase. This structure, referred to as a "reactive conflict graph," is a three dimensional geometric data container modeled after the reactive data structures of Oosterom.[28] The first two dimensions represent the geometry while the third dimension represents the scale.

**(2)** *Interactive Phase*. Once the reactive conflict graph has been populated, the interactive phase can begin. During this phase, the graph is queried for a specific map clipping and scale. The output of this query is a static conflict graph, which contains all the objects to be labeled and all possible conflicts between these objects given the user's current location and zoom scale. This process is illustrated in Figure 7.

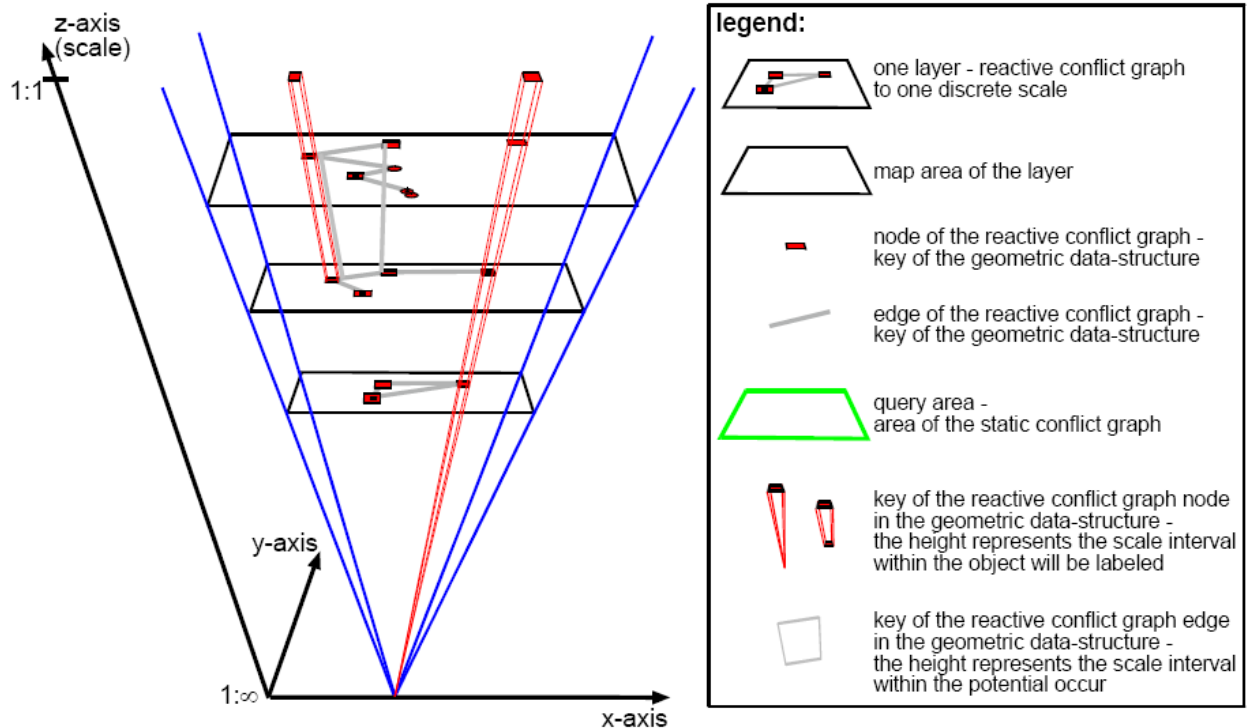

**Figure 7: A static conflict graph embedded in a three-dimensional reactive data structure.**

The prototype developed to demonstrate this approach does indeed yield millesecond responsiveness in the interactive phase. Moreover, this dynamic phase is not dependent upon any specific deconfliction algorithm. Any efficient heuristic algorithms, such as greedy, gradient descent, or simulated annealing, may be used for the final label selection. It should be noted that, although line and area labeling concepts are developed conceptually in Petzold's publications, complete implementations are only realized for point labeling problems.

### 2.5. MORE RECENT ADVANCES

In recent years a few notable advances have been made along a number of different lines. Some of these have yielded notable speed increases, and another has more precisely specified a formal model of dynamic maps, along with specifications for desirable interactive performance.

In response to Petzold's approach, Poon *et al.*[33] begin by observing that the former strategy (a) is too slow and (b) provides no guarantees of optimality at any given scale. To address these shortcomings they provide a carefully crafted response which is not only more efficient, but also provides theoretical guarantees. In so doing, they provide a methodology that permits "adaptive zooming" in such a way that information obtained from different resolutions is exploited to avoid having to recompute the entire labeling from scratch every time the zooming factor changes.

To achieve this they start by defining the zooming problem properly and precisely, and then they build a low-height hierarchy for efficient adaptive queries with theoretically guaranteed output. By observing that a zooming query in a window *W* is identical to a font-scaling query within *W* they are able to simplify notations throughout the paper.

The argument is built in two sections. The first section presents the adaptive zooming procedure in the one dimensional case. A greedy algorithm is utilized to select labels at the smallest or finest resolution (representing the state of being zoomed in far enough to disambiguate all the labels). A hierarchy of data structures is then consecutively built, each level of which represents one resolution or scale. Successive levels are constructed efficiently by observing that labels need only to be compared to the set of selected labels in the previous optimally arranged layout, thereby significantly reducing the required number of pairwise comparisons. The second section of the paper extends this line of argument to the two dimensional case. This achievement is enabled through the use of a technique known as "line-stabbing." This method, used also in [1], allows the one dimensional case to be extended in a straightforward way: horizontal lines, separated by the distance equal to the height of the labels are arranged so as to stab all the labels. The labels stabbed by alternating lines (all the even lines, or all the odd) are guaranteed not to overlap. Therefore, by performing the one dimensional algorithm on each line, and then choosing the maximal set of alternate lines (i.e., all odd or all even), a 2-approximation is easily obtained. Thus an $O(\log n)$-height hierarchy is built in $O(n^2)$ time and in $O(n \log n)$ space.

While their paper represents a solid algorithmic milestone, the achievement is principally theoretical. No prototype was implemented and no runtime measurements are offered. Indeed, although the guarantee of optimality may well be beneficial in some applications, it is unclear whether the $O(n^2)$ requirement will allow adequate responsiveness for interactive applications.

### 2.5.2. ROY'S "FAST" ALGORITHM

Just as Poon built upon the work and strategies of [1] so also Roy *et al.* [36] extend Agarwal's methodology, albeit in a different manner. In a paper which represents one of most efficient strategies to date, Roy outlines an approach which exploits the techniques of graph theory to solve a special case of the label maximization problem. The special class of the problem is restricted in the following ways: (1) labels are fixed height (but variable width), (2) the label must touch its referent point at one of the four corners, and (3) labels are not allowed to obscure other points. The authors describe a heuristic algorithm which runs in $O(n\sqrt{n})$, worst case. Their approach is favorably compared to the "Three Rules" approach, discussed above.

The technique is described in two phases: In the first phase a label graph is constructed which contains eight separate label collections. Each collections is guaranteed to represent an independent set (not necessarily maximal). The sets are generated as follows: for each point in the set of map features, four separate label configurations are established (as in Figure 8).

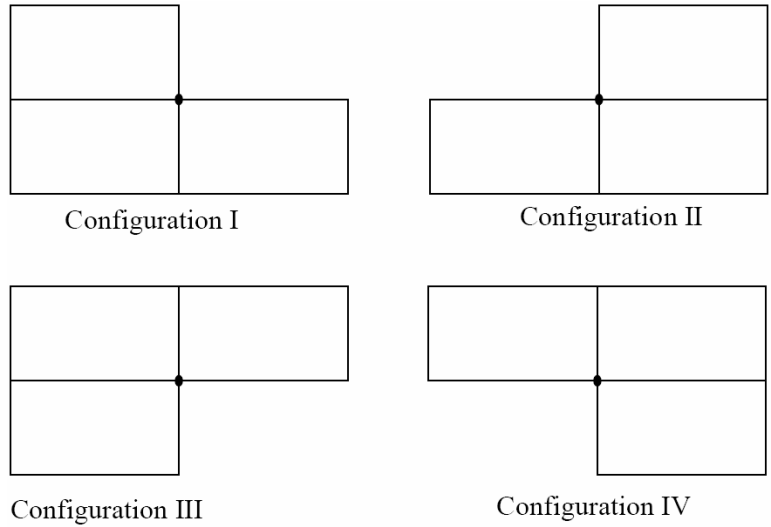

Configuration I         Configuration II

Configuration III       Configuration IV

**Figure 8: Possible configurations of valid labels, under Roy's algorithm**

After sorting the labels according to their $x$- and $y$-coordinates, alternating horizontal and vertical line sweeps are performed (two for each configuration), and thus, eight sets of non-overlapping labels are determined.

The second phase of the algorithm uses a bipartite graph, maximum independent set algorithm to merge the eight sets, two at a time, according to the schedule illustrated in Figure 9.

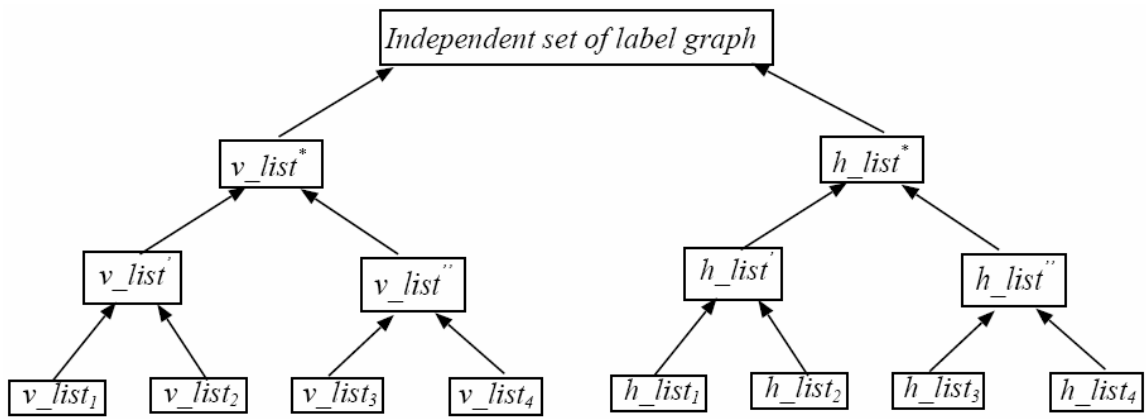

**Figure 9: Hierarchical merging in Phase 2 of Roy's algorithm.**

This merging operation requires the inspection of nine different ($h\_list*$, $v\_list*$) pairs to consider all the possible pairwise permutations. Thus, since each permutation requires seven executions of the merge algorithm, the technique requires 63 separate passes. Nevertheless, due to the fact that the algorithm operates in $O(n\sqrt{n})$ time, the authors are able to record very favorable time measurements.

Unlike the previous approach, these authors include a measured implementation of their approach. Using standard benchmark data they are able to demonstrate that their algorithm produces results equal or better to those presented in [43]. They point out that they are able to produce results quickly for a set of data that could not be processed in a day by the previous approach.

Due to these excellent results, this approach was the one chosen in Section 5 by which to compare the algorithmic approach proposed in the current paper.

### 2.5.3. BEEN'S DESIDERATA

The final paper to be considered is notable not only for its ability to achieve interactive speed, but even more so for the formal models it introduces. Been *et al.* in [2] have separated this problem into the two interrelated issues of label consistency and interactive speed. In the process they introduce the formal models of dynamic maps and dynamic labels. They formulate a general optimization problem for dynamic map labeling and give a solution to a simple version of the problem. The simple version is based on label priorities and a versatile and intuitive class of dynamic label placements they refer to as "invariant point placements." An appendix is also included in which they prove the *NP*-completeness of the simple dynamic labeling optimization problem. This paper builds significantly on the work of Petzold. In particular, these authors borrow from their predecessor the strategy of two-phase implementation: They separate the $O(n^2)$ task of determining the actual label placements from the final filtering stage. The former task is handled in a pre-processing phase, reserving the latter for the interactive phase.

Throughout this significant paper, several concepts are introduced and formalized, which can be summarized as follows:

(1)  ***Label placement consistency, or "Desiderata."*** Four invariant rules to guide the decision of label placement for a given point are presented. These so-called "desiderata" are assumed, but not defended, to be preferrable. They can be summarized as follows:

a.  Generally, labels should not vanish when zooming in or appear when zooming out.

b.  A label's position and size should change continuously under the pan and zoom operations (i.e., no "popping" or moving about unexpectedly).

c.  Labels should not vanish or appear during panning (until they slide completely out of view).

d.  The placement and selection of any label is a function of the current map state (scale and view area).

The authors point out that, to their knowledge, no previous approach has achieved these desiderata in implementation, although they cite other papers that recommend similar rules. In order to achieve such consistency their solution uses an "inverted sequence" by which all labels are first located, and then selected. These desiderata (denoted in their paper as (D1) – (D4)) inform all the design decisions which follow.

(2)  ***Invariant point placements.*** Key to their concept of consistency, particularly with respect to (D2) and (D4), is the idea that the label of a given point should not change through user interaction. On the other hand, the *size* of the label will indeed change (relative to the map, but not to the screen) as the user zooms. To capture these ideas, they model the dynamic point placements as truncated extrusions (cf., Figure 10(a)).

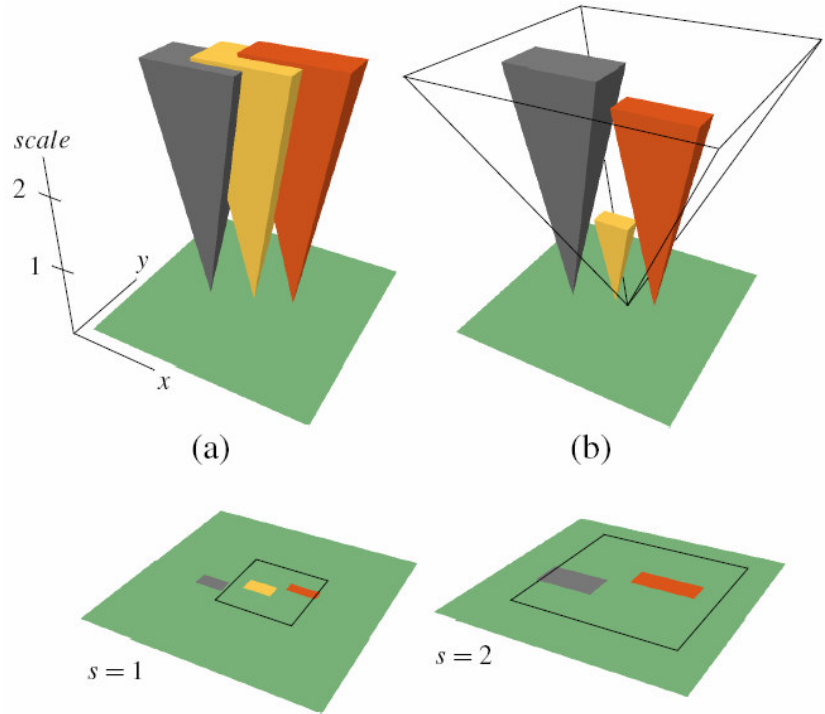

**Figure 10: Dynamic "invariant" placements for three labels, in Been's model.**

(3) ***Blocking scales.*** Following directly from the invariant point placements, the authors introduce the concept of "active" scales. Using point priority as a guideline, a given label's active scale begins at zero (thus adhering to (D1)), and extends just to the point where its extruded cone would overlap that of a higher priority label. At that top end of the active scale, the extrusion is truncated (cf. Figure 10(b)), thus eliminating that label from view at all higher scales.

(4) ***Implementation and practical refinements.*** Having thus formally defined the task at hand, the authors discuss their specific implementation. Significantly, they have successfully built a publicly available prototype that demonstrates the efficiency of their algorithm on a dataset of over 12 million points on a geographic map of the U.S. In order to accomplish this, a number of refinements were included. First, "levels of detail" (LOD) were established by which the dataset is segmented into eight non-overlapping regions of interest (i.e., street-level, city-level, state-level, etc). Each LOD is labeled independently of the others and inter-LOD coordination is not implemented. Second, the map is subdivided into a grid system of "buckets", which are exploited both to aid in geographic filtering, as well as to reduce the $O(n^2)$ time for computing all the active label scales. This latter use bears remarkable similarity to the independently devised "trellis" system that will be elaborated upon in this paper. The grid of buckets amount to a less-refined means to determine the neighboring points in a given region.

Indeed, as the publicly available prototype demonstrates, Been's approach clearly achieves seamless interactive performance speeds. The results of this achievement will undoubtedly shape the literature for years to come. Nevertheless, it is clear that this interactive responsiveness comes with the restrictive requirement of an intensive pre-processing phase. Unfortunately, for the purposes of the target problem described in this paper, such an encumbrance is not an option for dynamically generated datasets. In the narrower domain of geographic information systems, however, this approach clearly represents a watershed.

───────────────────────────────

As this brief overview of the historical literature illustrates, the history of this problem has evolved dynamically over the last twenty to thirty years. Advances have been made on many fronts, and clever new techniques and strategies continue to develop and mature. Every indication is that this process will continue for years to come. Indeed, critical breakthroughs seem imminent. Dr. Herbert Freeman, whose work in this field spans decades, offers an excellent synopsis of the last 25 years of progress on automated cartographic text placement systems. He concludes his overview with the promise that "the day of one-second quality labeling of an electronically displayed…map or…chart is not far off" [16].

## 3. STATEMENT OF PROBLEM

The goal of this research is to identify a method for labeling the features on dynamic maps in real time without a pre-processing stage. The term "maps" will be used in this paper to refer to any chart, view, or diagram produced by information visualization and visual analytic software, in which the main features (i.e., elements to be labeled) are represented as *points*. These "point-features" will be referred to henceforth as simply "features". The set of all features will be called the map set. These maps are considered "ad hoc" in that they are generated on the fly in unpredictable configurations and require a labeling method that can be calculated instantly. Moreover, the data sets can be massive: ranging up to tens of thousands of features or more.

The precise scope of the problem can be defined as follows: Given a map with a pre-defined set of prioritized features, swiftly find the largest possible set of non-intersecting, axis-aligned rectangular labels, giving preference to the highest-prioritized features and to the cartographically preferred candidate locations. Each label must: (a) touch its referent (parent) feature at one of its four corners, (b) obscure as few other features as possible, and (c) obscure no other labels.

In seeking to address this problem, the following design decisions have been made in the algorithm to address the specific challenges posed by dense feature-sets in interactive displays. Although they are somewhat restrictive, the implementation described in this paper provides compelling evidence that this approach can yield highly effective and useful results.

***It does not restrict labels from obscuring other features in the view.*** This approach represents a significant departure from most previous schemes, but it is clearly unavoidable when working with the dense clusters produced by visualizations of massive data sets. The on-screen density of these visualizations often provides little or no white space for non-conflicted labels.

To prohibit occlusion of features in these cases would mean that very few features (if any) would be labeled at all. Worse, because the highest-prioritized features are often buried deep in clusters, only the marginal outliers would be labeled, in direct opposition to our stated goal of priority preference. That being said, this algorithm attempts to allow the fewest features possible to be obscured, and then, only by higher priority feature labels.

**It is optimized for uniformly-sized labels.** The trellis strategy, as described in Section 4.1 below, is designed primarily for labels of equal size. Nevertheless, non-uniform labels can be accommodated, albeit with diminishing quality. For example, users could specify that longer-than-average labels may be truncated, or alternatively, may be allowed to overlap other labels by a certain percentage. Experimentation has demonstrated that a combination of these two options can produce very acceptable and readable labelings with variably sized labels.

**It exploits priority-ranked features.** It is common in visual analytic applications that the individual features are associated with a specific value representing their "importance" to the view. This value may be derived or assigned. For example, if the view represents a geographic map with features pinpointing the cities, the features might be ranked according to the population of those cities. Or if the view was a complex scatter-plot displaying the results from a web search, the feature priority might correspond to their weight, Google™ PageRank, chronological order, or any other ordered numeric attribute of the data. The use of such preference information, if available, can greatly increase the efficiency of the algorithm. See [2] for similar use of priority rankings. Prioritized or "weighted" feature data has also been exploited in [32] and [29].

**It prefers speed over optimal label configurations.** This design decision is based squarely on the interactive nature of dynamic maps. The assumption here is that any sub-optimal configuration or indistinct labeling can in most cases be disambiguated with a minor amount of user interaction (e.g., zooming). While this allows us to relax our expectations slightly with regard to perfect labelings, experimental results demonstrate that quality is only marginally degraded.

***It is based on a label model of size four***. A "label model" refers to *the number of possible positions,* or "*candidates,*" in which a label may be located around a given feature. Many mapping utilities and geographic information systems use label models offering more than four candidates per feature. Some offer the "slider" model, where labels are allowed to "slide" around the feature so that the feature may appear anywhere on the boundary of the label (cf. [41, 49]). However, the contention here is that such extended label models are inappropriate for dense visualizations. Given the magnitude of screen-space density in high-volume visualizations, particularly when labels are allowed to over-post features as discussed above, referent-ambiguity increases uncontrollably as the label model grows. Because so many features may feasibly appear on or near the boundary of a label, only by constraining the labels so that their referent feature appears at one of their corners can we preserve appropriate contextual awareness (cf. Figure 17). Similarly, the "excentric" model, where labels are attached by leader lines to their associated features [3, 14], is not particularly suited for high-density displays.

***It recognizes cartographic preference***. The relative preference of each of a label's four candidates is determined by the aesthetic criteria established by cartographers. Described in [21], these informal rules establish the relative value of each candidate. Typically, the upper-right candidate is preferred, followed by the lower-right, upper-left, and lower-left, in that order (cf. Figure 11). Other aesthetic issues are discussed and evaluated in [20].

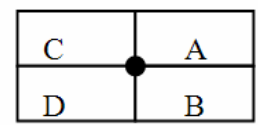

**Figure 11: Cartographic preference of label candidates around a feature point.**

# 4. ALGORITHM: LABEL SELECTION BY CONFLICT- EXPENSE ESTIMATION

This section presents a solution to the stated problem, introducing a rapid, yet effective method of determining and selecting what will be referred to as the "least expensive" label configuration. The algorithm is divided into the following three conceptually distinct steps:

- Conflict detection

- Expense calculation

- Label candidate selection

In implementation, these steps would actually be merged. *First*, analyze each feature in the set, testing it against all features that lie in close proximity, and create what is known as a "conflict graph" to keep a record of label candidate intersections, or "conflict partners". Although this step would normally be quite expensive, an efficient approach—the trellis strategy—will be introduced to expedite these calculations. *Second*, pass through the set of label candidates and, for each one, calculate an associated *expense*. This expense will be described in detail below but is essentially a function of the sum of the label's conflict partners along with their associated priorities, among a few other factors. *Finally*, pass through the set of features, in descending priority order, and for each feature, select its "least-expensive" label candidate, while de-selecting that feature's remaining candidates. In essence, this algorithm uses an informed but greedy approach to approximately minimize the expense of the total label set.

By computing each candidate in priority order, we can have high confidence that a large majority of the highest priority features will be labeled at the default zoom level, assuming an approximately uniform distribution of the visualization data features. This system does not offer demonstrable asymptotic guarantees of performance, as it claims only to be a useful approximation. Nevertheless, the author's implementation provides strong evidence that this approach is fast, effective, and reliable in many real-world scenarios. Furthermore, it should be reiterated that this system requires no extended pre-processing stage, making it ideal for information visualization and visual analytic applications in which maps, diagrams, and charts are constructed on the fly. This is particularly true of three-dimensional feature maps, for which preprocessing every conceivable orientation is entirely unfeasible.

### 4.1. STEP ONE: CONFLICT DETECTION

One of the most common approaches in the literature to solving the point-label problem uses a concept referred to as a "conflict graph [39]", an "overlap graph" [23], or a "label graph" [36]. These terms refer to a graph whose nodes correspond to labels and whose edges correspond to intersections between labels. Using this model, the objective of the label placement problem is to find the maximum independent set of the conflict graph [1]. The concept of a conflict graph can also be understood more generally, in a non-graph-theoretic sense, as an "adjacency matrix" storing intersection information for all feature pairs. This paper generalizes the concept of a conflict graph, defining it simply as a list of conflict partners associated with each feature. This list will act as a look-up table to determine the total number of conflicting pairs of label candidates or, more specifically, the total number of occluded labels for any given candidate, and to keep record of the label candidates that must be removed due to occlusion.

Although constructing the conflict graph is typically quite expensive, this step is often disregarded in the literature. The conflict graph is quite often assumed as input to many algorithms (see for example [2, 43, 46]). Such papers generally describe how to determine or identify the largest independent set of label candidates, given a *pre*-determined conflict graph. This approach is appropriate if the size of the dataset is relatively small, or if the pre-processing time available for conflict graph construction is irrelevant. Unfortunately, neither of these assumptions holds true in the realm of dynamic information visualization applications. Indeed, even though many of these approaches are offered for their efficiency, this overlooked step can often be the bottleneck of performance.

In fact, some more recent papers have sought greater efficiency in label-selection by *intensifying* this requirement, rather than abbreviating it. In [29, 30], for example, an approach is described for determining a "reactive conflict graph." This graph stores information about all potential conflicts at all zoom levels. It is constructed during a several-minute-long "pre-processing" phase and rapidly accessed during an interactive phase. In [33], a line-stabbing approach to conflict graph generation is used, building low-height hierarchies to better ensure the quality of the final layout at multiple zoom levels. This approach has the advantage of offering asymptotic guarantees of performance but requires $O(n^2)$ time for two-dimensional maps. Likewise, interactive speed is achieved in [2], but "all of the selection and placement decisions are moved into the preprocessing stage." These approaches, with their reliance upon a time-consuming pre-processing stage, make them unsuitable options for the ad hoc map generation of dense information visualization applications.

Conceivably, one could attempt to eliminate conflict graph construction altogether by simply estimating the number of conflicts for any given feature. By using, for example, a modification of the "trellis" approach described in this paper such estimates could indeed turn out to be reliable and sufficient. However, such an approach does not address the second purpose of the conflict graph mentioned above; that is, it provides no way to determine which specific labels must be removed due to occlusion. For this reason, a conflict graph of some form must be generated and must be available and accurate at all zoom-levels and orientations of the feature set although the construction of a *globally* complete graph is not feasible for any fast de-conflicton algorithm, the ideal solution must at least include a strongly reliable *approximation* of the graph. This necessity will require a considerably more sophisticated and informed approach to conflict graph generation than the naïve $O(n^2)$ method of testing all candidates against all others. The following section will describe in detail a strategy that can in principle offer an efficiency gain of nearly three orders of magnitude over the elementary approach. The dual goal of this new approach is to accelerate the detection of intersecting label candidates, and to avoid all unnecessary testing of feature pairs.

### 4.1.1. THE TRELLIS STRATEGY

The strategy employed here is based on a geometric subdivision of the feature set in screen-space. This subdivision, called a "trellis," can be conceived as a two-dimensional array of equally sized cells, sub-dividing the screen-space view into rows and columns. Each cell in the trellis has an associated list of all the features that are located in that cell's defined pixel space. The trellis, of course, would never actually be rendered in any visualization, but if it were it would resemble a cross-hatch lattice of horizontal and vertical lines—like a garden trellis. This trellis is similar in many respects to the "grid of buckets" described in [2]. The approach described here, however, exploits the characteristics of a slightly more sophisticated subdivision of map space.

One very significant and strategically determined characteristic of the trellis is the size and aspect ratio of the cells. Each cell is defined as a "*quarter-region*"; that is, it is exactly one-fourth the area associated with a label, the size of which we assume to be fixed and constant (cf. Section 3). It also shares the same aspect ratio of the label. In other words, if one were to slice a label in half both vertically and horizontally, the resulting four quarter regions define the trellis cell (cf. Figure 12). (Henceforth, "trellis cell" and "quarter-region" will be used interchangeably.) The purpose for this particular size and shape will be explained shortly. All the cells in the trellis are, hence, identical, with the possible exception of the rightmost column and the bottom-most row of cells, which may be smaller if view-space constrains them. For a view of size 1500x1000 pixels, with labels of size 150x20, there would be 2000 cells in the trellis.

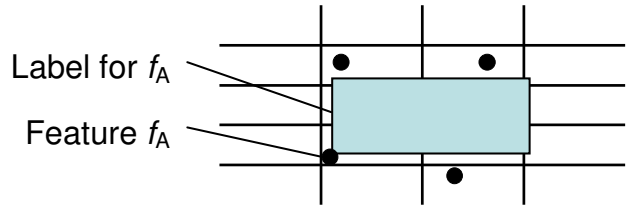

**Figure 12: A portion of the trellis. The label of feature point $f_A$ is displayed to demonstrate that each trellis cell is one-fourth the size of the labels. Other features are shown in the neighborhood of $f_A$.**

The trellis is populated by testing each feature $f$, performing the following two calculations to determine which row and column of the trellis contains the feature.

$$\text{Trellis}_{\text{column}}(f) = \text{floor}(f_x/\text{quarter\_region\_width})$$
$$\text{Trellis}_{\text{row}}(f) = \text{floor}(f_y/\text{quarter\_region\_height})$$

where $f_x$ and $f_y$ represent the x and y pixel coordinates of $f$, respectively, relative to the view window. Hereafter the trellis cell in which feature $f_A$ lies will be referred to as CELL($f_A$). The expense of this series of calculations is, of course, linear in $n$.

Once we have iterated the feature set, every trellis cell is associated with a list of the features that appear within its boundaries. Conversely, every feature has an associated trellis coordinate. By taking advantage of the specific geometric structure and layout of the trellis, the conflict graph can be generated in an extremely efficient way.

### 4.1.2. THE TRELLIS ADVANTAGE

The first and most obvious advantage offered by the trellis is the ability to specifically and instantly define a neighborhood around any given feature. As a result, we will not have to test each feature against every other feature in the set (an $n^2$ operation). Instead, we can limit our conflict search so that, for any given feature, we will test it only against those features that are located in the immediately neighboring cells. Specifically, we will inspect only those cells that have a possibility of containing conflicting label candidates. This approach was similarly described in [2].

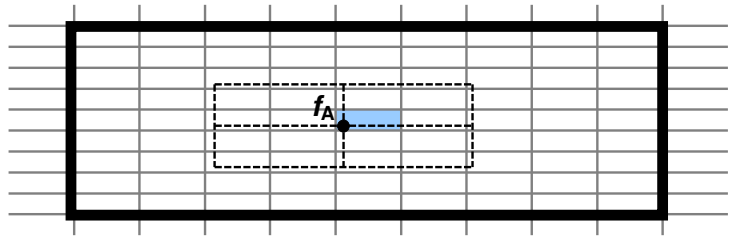

**Figure 13: A portion of the trellis displaying the entire neighborhood(bold line) of feature $f_A$. Four candidate label locations are shown (dashed lines). Note that regardless of where $f_A$ is located in its parent cell (blue), it is guaranteed not to conflict with the labels of any features outside this neighborhood.**By the nature of the design of the trellis dimensions, it is a simple task to determine which cells have potential conflicts. Because quarter-regions are ¼ the size of a label, it may be observed that a label candidate of feature $f_A$ can conflict with a label candidate of feature $f_B$ if and only if $f_B$ resides in a four-cell radius of $f_A$, or no more than four rows or four columns away from $f_A$. Such a radius defines a 9x9 array of cells, centering on the home, or "parent" cell of $f_A$ (cf. Figure 13). This array will hereafter be referred to as the "neighborhood" of $f_A$, and features within those cells, as "neighboring features" of $f_A$. All features outside of this neighborhood can be disregarded with respect to $f_A$; their label candidates have no possibility of conflicting with those of $f_A$ beyond a coincident edge. On the other hand,

the label candidates of all neighboring features of $f_A$ are regarded as potential conflict partners with the candidates of $f_A$.

Referring to the example view mentioned above, we can see that this has reduced the target range of test cells from 2000, which was the number of cells in the entire view, to 81. If at least a nominally uniform distribution of features is assumed, it can be seen that this strategy has reduced the number of potential conflict tests by roughly one order of magnitude.

The benefit of subdividing our space into quarter-regions is that it can significantly reduce the number of conflict tests necessary. For instance, consider a feature $f_A$ in CELL($f_A$), and a second feature $f_B$ in one of the neighboring cells. Because CELL($f_B$) is in the neighborhood of CELL($f_A$), their label candidates are potentially in conflict. Imagine next that CELL($f_B$) is exactly four cells to the left of CELL($f_A$). Due to this fixed Cartesian relationship we can narrow the list of potential conflicts, thus eliminating the need to test all of them. The complete list of possible conflicts between the label candidates of $f_A$ and $f_B$ in this case are as follows:

[A0:B2, A0:B3, A1:B2, A1:B3]

We therefore only have to test those particular candidates for conflicts rather than all 16 possible conflict pairs. Tables 1-3 delineate precisely all the possible conflicts that can occur in each cell in the neighborhood of a central feature. Moreover, as we continue to exploit the known Cartesian relationships between the cells, we gain an additional advantage. Note that, typically, conflict graph generation requires the detection of rectangle intersections, which is not necessarily an inexpensive test. Many approaches to the general problem of rectangle intersection-pair identification have been published within the field of computational geometry (e.g., [7, 18]), but the problem itself is a difficult one. Even the simple case of testing a single pair of rectangles is fairly incompressible, with essentially no way to abridge it beyond four atomic operations, such as

```
IntersectionExists(rectA, rectB) =
((rectA.left < rectB.right) && (rectA.right > rectB.left)
    && (rectA.top > rectB.bottom) && (rectA.bottom < rectB.top));
```

Fortunately, however, the use of the trellis method eliminates the need to perform any rectangle intersection tests at all. Consider the example described above, in which $f_B$ lies 4 cells to the left of $f_A$, and four possible candidate intersections exist between them. Rather than testing each one of the four pairs for intersection, we will exploit the information provided to us by the geometric orientation within the trellis. In this particular case, then, we merely have to test whether the distance from $f_A$ to $f_B$ is greater than the width of two labels, in which case no conflicts are possible. If the distance is less than two label-widths then we additionally test whether $f_A$ is higher (in the $y$-direction) than $f_B$. The result of these two tests determines precisely which of the four possible label conflicts actually occur.

We can use similar observations in each cell of a given neighborhood to greatly reduce the number of required tests. Table 1 represents one quadrant of the trellis neighborhood and outlines all the tests required for each cell in that area. (The complete table is reproduced in Appendix B). As the table demonstrates, none of the cells in the entire 81-cell neighborhood require more than two atomic tests; many require only one; and a few require none at all. If one feature were to reside in every cell of the neighborhood of $f_A$ there would then be at most 90 tests required to build the complete conflict graph with respect to $f_A$. Note that the simplest rectangle intersection approach would require testing all four of $f_A$'s label candidates against the four candidates of each of its 80 neighbors, or 4x4x80=1280 total tests. If the cost of the four-operation intersection test were included, the total count would approach 5000 atomic operations.

Contrast this with the 90 operations of the trellis strategy, and it can be seen that the efficiency has been increased by a factor of nearly 500:1. The trellis strategy has, hence, reduced the cost of the problem by a second order of magnitude, for the uniformly-distributed case.

This demonstrates the tremendous benefit offered by the trellis method. For the low ($O(n)$) cost of a single pass through $F$ to populate the trellis, a potential speed-up of nearly 2½ orders of magnitude has been achieved over a rudimentary approach to conflict graph generation. In addition, another characteristic of the trellis will be used shortly in calculating the values of each label candidate.

| 1  -4/-4 | 2  -3/-4 | 3  -2/-4 | 4  -1/-4 | 5  0/-4 |
|---|---|---|---|---|
| $\alpha0, \alpha1, \beta1$ | $\alpha1, \beta1$ | $\alpha1, \beta1, \gamma13$ | $\alpha1, \gamma13$ | $\alpha1, \gamma13, \gamma31$ |
| ΔY>2? <br> / \ <br> $\alpha0$ : ΔX>2? <br> / \ <br> $\alpha1$ : $\beta1$ | ΔY>2? <br> / \ <br> $\alpha1$ : $\beta1$ | ΔY>2? <br> / \ <br> $\alpha1$ : ΔX>1? <br> / \ <br> $\beta1$ : $\gamma13$ | ΔY>2? <br> / \ <br> $\alpha1$ : $\gamma13$ | ΔY>2? <br> / \ <br> $\alpha1$ : $X_A$>$X_B$? <br> / \ <br> $\gamma13$ : $\gamma31$ |
| **10**  -4/-3 | **11**  -3/-3 | **12**  -2/-3 | **13**  -1/-3 | **14**  0/-3 |
| $\alpha0, \beta1$ | $\beta1$ | $\beta1, \gamma13$ | $\gamma13$ | $\gamma13, \gamma31$ |
| ΔX>2? <br> / \ <br> $\alpha0$ : $\beta1$ | No test | ΔX>1? <br> / \ <br> $\beta1$ : $\gamma13$ | No test | $X_A$>$X_B$? <br> / \ <br> $\gamma13$ : $\gamma31$ |
| **19**  -4/-2 | **20**  -3/-2 | **21**  -2/-2 | **22**  -1/-2 | **23**  0/-2 |
| $\alpha0, \beta1, \gamma10$ | $\beta1, \gamma10$ | $\beta1, \gamma10, \gamma13, \delta1$ | $\gamma13, \delta1$ | $\gamma13, \gamma31, \delta1, \delta3$ |
| ΔX>2? <br> / \ <br> $\alpha0$ : ΔY>1? <br> / \ <br> $\beta1$ : $\gamma10$ | ΔY>1? <br> / \ <br> $\beta1$ : $\gamma10$ | ΔX>1? <br> / \ <br> ΔY>1? : ΔY>1? <br> / \ <br> $\beta1$ : $\gamma10$   $\gamma13$ : $\delta1$ | ΔY>1? <br> / \ <br> $\gamma13$ : $\delta1$ | ΔY>1? <br> / \ <br> $X_A$>$X_B$? : $X_A$>$X_B$? <br> / \ <br> $\gamma13$ : $\gamma31$   $\delta1$ : $\delta3$ |
| **28**  -4/-1 | **29**  -3/-1 | **30**  -2/-1 | **31**  -1/-1 | **32**  0/-1 |
| $\alpha0, \gamma10$ | $\gamma10$ | $\gamma10, \delta1$ | $\delta1$ | $\delta1, \delta3$ |
| ΔX>2? <br> / \ <br> $\alpha0$ : $\gamma10$ | No test | ΔX>1? <br> / \ <br> $\gamma10$ : $\delta1$ | No test | $X_A$>$X_B$? <br> / \ <br> $\delta1$ : $\delta3$ |
| **37**  -4/0 | **38**  -3/0 | **39**  -2/0 | **40**  -1/0 | **41**  0/0 |
| $\alpha0, \gamma10, \gamma01$ | $\gamma01, \gamma10$ | $\gamma01, \gamma10, \delta0, \delta1$ | $\delta0, \delta1$ | $\delta0, \delta1, \delta2, \delta3$ |
| ΔX>2? <br> / \ <br> $\alpha0$ : $Y_A$>$Y_B$? <br> / \ <br> $\gamma01$ : $\gamma10$ | $Y_A$>$Y_B$? <br> / \ <br> $\gamma01$ : $\gamma10$ | ΔX>1? <br> / \ <br> $Y_A$>$Y_B$? : $Y_A$>$Y_B$? <br> / \ <br> $\gamma01$ : $\gamma10$   $\delta2$ : $\delta3$ | $Y_A$>$Y_B$? <br> / \ <br> $\delta0$ : $\delta1$ | $X_A$>$X_B$? <br> / \ <br> $Y_A$>$Y_B$? : $Y_A$>$Y_B$? <br> / \ <br> $\delta0$ : $\delta1$   $\delta2$ : $\delta3$ |

**Table 1: The Trellis Neighborhood Test Outline** (showing only the upper-left quadrant of the neighborhood, due to space constraints). The central cell is bottom right (in blue). The other three quadrants (not shown), are symmetric, but not identical, to this one. Each cell is defined as a "quarter-region" being one fourth the size of an actual label. The specific tests, in bold, are explained more fully in Figure 15. The result of each test is a particular label conflict configuration, abbreviated with codes such as $\alpha1$, or $\beta1$, and delineated precisely in Table 2.

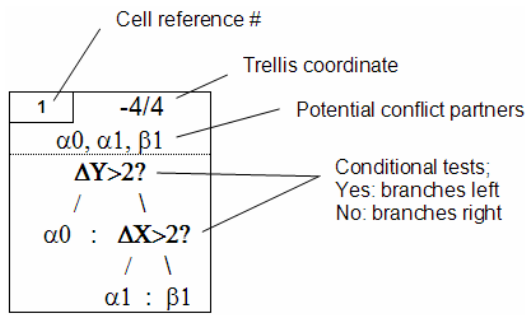

**Figure 14: Legend for Test Outline**

| **Conditional Tests:** | |
|---|---|
| ΔY>1? | → abs($f_A$.y_coord - $f_B$.y_coord) > Label_height |
| ΔY>2? | → abs($f_A$.y_coord - $f_B$.y_coord) > 2*Label_height |
| $Y_A$>$Y_B$? | → $f_A$.y_coord > $f_B$.y_coord |
| ΔX>1? | → abs($f_A$.x_coord - $f_B$.x_coord) > Label_width |
| ΔX>2? | → abs($f_A$.x_coord - $f_B$.x_coord) > 2*Label_width |
| $X_A$>$X_B$? | → $f_A$.x_coord > $f_B$.x_coord |

**Figure 15: A complete list of the tests required by the Test Outline**

| Types: | Possible Label Pair Configurations | | | |
|---|---|---|---|---|
| **α** | α0 | α1 | α2 | α3 |
| CPs: | Ø | Ø | Ø | Ø |
| **β** | β0 | β1 | β2 | β3 |
| CPs: | A0:B3 | A1:B2 | A2:B1 | A3:B0 |
| **γ** | γ10 | γ13 | γ31 | γ32 |
| | A0:B2, A1:B2, A1:B3 | A1:B0, A1:B2, A3:B2 | A1:B0, A3:B0, A3:B2 | A2:B0, A3:B0, A3:B1 |
| | γ01 | γ02 | γ20 | γ23 |
| CPs: | A0:B2 A0:B3 A1:B3 | A0:B1 A0:B3 A2:B3 | A0:B1, A2:B1, A2:B3 | A2:B0, A2:B1, A3:B1 |
| **δ** | δ0 | δ1 | δ2 | δ3 |
| CPs: A0:B0, A1:B1, A2:B2, A3:B3 | A0:B1, A1:B3, A0:B2, A2:B3, A0:B3 | A1:B3, A3:B2, A1:B2, A0:B2, A1:B0 | A2:B0, A3:B1, A2:B1, A0:B1, A2:B3 | A3:B0, A1:B0, A3:B1, A2:B0, A3:B2 |

**Table 2: Label Pair Configurations.** This table represents every possible configuration between a given pair of features, along with the corresponding "conflict pairs" (CPs) among two label candidates, A and B. (Feature $f_B$'s candidates are smaller for illustration only). The notation [A2:B0] denotes: "$f_A$'s candidate #2 conflicts with $f_B$'s candidate #0." The configurations are grouped in four types (α, β, γ, δ). In α−configurations no conflict occurs between label candidates, whereas in δ−configurations all four of $f_B$'s candidates are obscured.

### 4.2. STEP TWO: COST ANALYSIS

Having established a conflict graph for each feature, the objective of the second step of the algorithm is to determine the least expensive label candidate position for each feature, among its four options. Each label candidate has an inherent value based on the priority of its referent feature. (Recall that the features are prioritized or ranked in order of preference.) The goal of the algorithm is simply to maximize the sum of all the values in a set of non-occluded candidates.

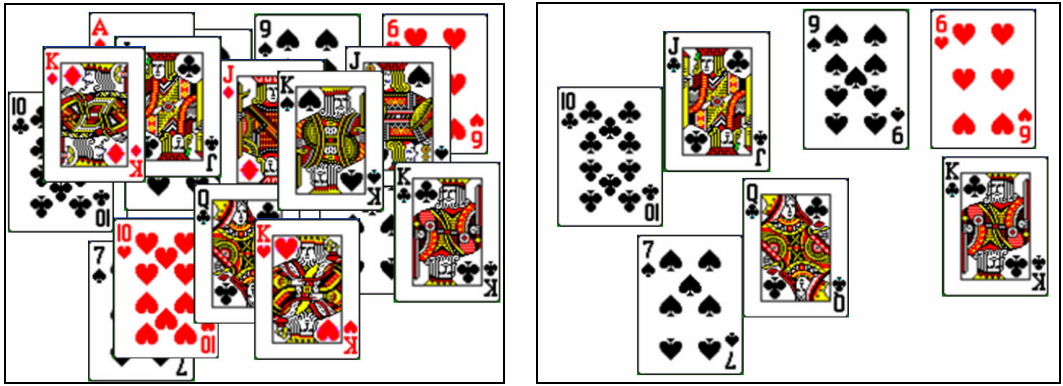

**Figure 16: An assortment of cards illustrating the "least expensive" deconfliction routine.**

By way of illustration, imagine a deck of playing cards randomly scattered face-up on a small table (cf. Figure 16). The face value of the card represents its priority and the suits represent the four label candidates of a given feature. Cards resting on top of other cards represent conflicting label positions. Assume that the cards were dealt in sorted order, so the kings are on top. The task is to choose at most one suit for each card value, while removing all cards that are partially obscured by higher valued cards. Once a suit is selected for a given card value all cards it "conflicts" with (the ones directly under it) must be removed from the table. The goal is to produce a set of non-occluded cards with the highest possible total value. By using this analogy it is fairly easy to see that one effective strategy would be to begin with the four kings and select the one which rests on the "least expensive" pile of cards. We call this selected card the least expensive candidate. Once this selection is made the other three kings are removed, and also all the cards that were under our selected king. We can then proceed in the same way with the remaining queens and so on. Determining which stack of conflicting cards is least expensive is simply a matter of finding the sum of the face value of those cards. This card game illustrates the strategy used by the algorithm of this paper. The goal, simply put, is to approximately maximize the set of all the values of non-occluded feature labels.

It should be noted that, although we have spoken of maximizing the expense of the mapset, this approach is not strictly an optimization algorithm. It is not claimed that the global maximum of the feature set expense will be found. The algorithm is essentially a "greedy" one and is indeed vulnerable to local maxima. Nevertheless, as will be demonstrated, this approach will provide a useful and reliable approximation of the optimal outcome. Our confidence in this approximation stems from the fact that we are processing our features in priority order. As we pass through our feature set in descending order, choosing the least expensive candidate for each feature, our guarantee is that no feature will remain unlabeled unless all of its candidates are occluded by *higher* priority labels. This approach is, of course, dependent upon an accurate assessment of the inherent value of each label. This value will now be examined more precisely, looking both at the static base value of a given label and, subsequently, at some dynamic runtime modifications to that value.

### 4.2.1. LABEL CANDIDATE BASE VALUE

As has already been noted, the default value of a given label is dependent directly on the priority of that label's referent feature. High priority features will obviously have more valuable labels. The model is further extended to include the "cartographic criteria" (described in Section 3) by assigning a higher value to the aesthetically preferred label candidates.

One drawback of this simplistic cost model is that it allows the possibility that a small cluster of low priority labels might "outweigh" a few higher priority labels. To return to our index card analogy, imagine that one top-card happens to be covering a large stack of low-ranked cards, while an adjacent top-card is overlapping only a few cards, but they are high-ranked. The algorithm would consider the first top-card to be the most expensive (if the sum of the numbers on the under-lying cards exceeds that of the smaller stack); that card would therefore be removed to reveal the cards beneath it, while the second top-card would be "selected" (causing the high-ranking cards below it to be removed). Furthermore, the "expensive" stack may well be so mutually conflicted that very few of them will remain on the table after all conflict-partners are removed. The subsequent "score", therefore would be lower than it would have been if the second top-card had been removed instead.[1] To be more specific, consider the case where feature $f_A$ with two label candidates, $f_A{}^0$ and $f_A{}^1$, where $f_A{}^0$ conflicts with a high priority feature $f_B$ and $f_A{}^1$ conflicts with a group of low priority features. Ideally, $f_A{}^1$ should be selected because it obscures only low priority features. However, because the algorithm sums the value of the conflicts, $f_A{}^1$

---

[1] One possible "fix" for this problem is based on the observation that the maximum number of non-conflicting labels that a given label candidate can occlude is four. [For a proof of this, assume to the contrary that a label could occlude five equally-sized, non-conflicted neighbor labels. (Recall our apriori assumption of fixed/constant label sizes.) Geometrically this would require a minimum of three of these neighbor labels to be positioned adjacent to each other (but, of course, not overlapping), and aligned either horizontally or vertically. In order for a label candidate to occlude all three of these adjacent neighbor labels, the second (middle) one in the group of three would have to be smaller in size than the occluding label (narrower if they are aligned horizontally, or shorter if they are aligned vertically). But this violates our apriori assumption of equally-sized labels. Hence, we have a proof by contradiction.] Since, then, a given label can occlude no more than four non-conflicting neighbors, an accurate cost analysis *only has to take into account the expense of the four highest ranking conflict partners for any given label candidate.* The cost of those four represents the maximum total "score" that could be obtained by removing an over-lapping label candidate.

The drawback of this solution, however, is the added performance cost of tracking the four highest ranking conflict partners. Doing this requires the system to maintain a sorted list of the top four partners of each candidate throughout the trellis neighborhood iteration. Although it is only the first four conflict partners in the list that need to remain in sorted order, this nonetheless adds multiple operations to every step of the algorithm. It is left to the implementer to decide whether the enhanced results are worth the added algorithmic expense. In light of the added expense of this process, therefore, an alternative (and much cheaper) solution will now be proposed.

may in fact be deemed more expensive, and consequently the high priority conflict partner $f_B$ would be occluded. To mitigate against circumstances such as this, the cost model requires the following adjustment. Taking $\text{VALUE}(f_0) = \text{BASE\_VALUE}(f_0)$, iteratively update the value of all remaining features like so:

$$\text{VALUE}(f_i) = \text{VALUE}(f_{i-1}) + \frac{\text{BASE\_VALUE}(f_i)}{n}, i = \{1, ..., n\}$$

Thus, the increment between any two consecutively prioritized features is changed from 1 to $i/n$ (where $\text{BASE\_VALUE}(f_0) \approx i$). This effectively spreads the data to give increasingly greater weight to the higher priority features requiring, for example, twice as many mid priority label candidates to "outweigh" a high priority candidate.

### 4.2.2. LABEL CANDIDATE VALUE MODIFICATIONS

Up to this point, the value of a given label candidate has been fixed— determined at the time of view-initialization as a function of its referent feature priority and its label-model (aesthetic) preference. This value has so far been independent of the conflict graph. In order to increase the effectiveness of the cost-analysis approach, however, two specific modifications will now be introduced to adjust these values dynamically, in response to circumstances detected or produced in the conflict graph construction phase.

First, the value of a candidate will be adjusted upwards each time one of its siblings is occluded. The notion behind this is that as a feature loses its candidate labels due to occlusion, the remaining candidates become increasingly valuable. If a feature has only one non-occluded candidate left, it should be considerably more expensive to de-select it. We therefore increase the value of each label by an amount equal to the value of the occluded siblings. Hence, the value of occluding a neighbor's final candidate is equal to occluding all four of that neighbor's candidates because both cases result in the de-selection of an entire feature, preventing that feature from receiving a label. Therefore, the value of an "only child" should be equal to the sum of all the original candidates. By extending this logic, the value of any label should be incremented by the value of each lost sibling.

The second dynamic modification to our cost analysis addresses the interactive nature of the map. The notion here is that, given the user's ability to zoom, many label candidates that are occluded at one particular zoom level will be freed up as the user zooms in closer. Therefore, many of the features in the distant cells of the neighborhood of $f_A$ will no longer remain in the neighborhood at deeper zoom levels. Because of this, the value of a conflict partner of $f_A$ should increase with its *proximity* to $f_A$. In other words, features that are in the nearest cells to $f_A$ would potentially require more zoom steps to de-conflict, and should therefore be more expensive to occlude. The exact amount of the value adjustment in most cases should be a function of the magnification factor applied to each level of zoom. The author's own implementation uses the following adjustment schedule for the modified value:

$$\texttt{MODIFIED\_VALUE}(f_B) = \texttt{BASE\_VALUE}(f_B) * (\texttt{PROX\_WT}*(5-\texttt{RAD\_DIST}(f_A, f_B)))$$

where PROX_WT is the proximity weight factor (in this implementation it is .5), and RAD_DIST is the radial distance from $f_A$ to $f_B$ in cells (which ranges from 0 to 4). This distance is counted in rows or columns of separation, whichever is greater. This divides the neighborhood into four concentric rectangles around the parent cell of $f_A$ and increments the value between each progressively closer ring of cells by 50%. The ultimate effect of this adjustment is that the algorithm can "intuitively" determine the least expensive labels over several zoom levels. This will tend to mitigate inconsistency in label candidate choice from one zoom level to another.

### 4.3. STEP THREE: LABEL SELECTION

Once a weighted set of label candidates has been determined for each feature, the final label selection can be made by simply comparing the non-occluded candidates of each feature and selecting the least expensive one. All other candidates will be de-selected, which essentially disposes of them for the purposes of this algorithm, until the next viewer interaction reinitiates this entire process. Finally, using the conflict graph built in the first step, all conflict partners of the selected label are de-selected by occlusion. They are thereby removed from their respective referent feature's pool of label candidates.

Two additional refinements increase the algorithm's efficiency even further:

First, any feature $f_A$ that has already lost all of its candidates to occlusion from higher priority labels need not be tested against its neighbors. Doing so would be a waste of cycles, as none of $f_A$'s labels have any chance of being selected. In particularly dense sets, this can automatically prune a significant number of features.

Second, as a feature is being tested against its neighbors, it can safely ignore all higher priority features as it is guaranteed not to have any conflicts with them. If a conflict had existed in the initial configuration between a candidate of feature $f_B$ and that of a higher priority feature $f_A$, it would have already been detected during $f_A$'s neighborhood traversal. In that case, either the conflicting candidate of $f_A$ was selected, and the occluded candidate of $f_B$ removed, or else a different candidate was selected for $f_A$ and the conflicting candidate was removed from contention. In both cases, the conflict is removed prior to the testing of $f_B$. Using this shortcut can cut the required number of tests in half.[2]

As was stated previously, this algorithm was divided, for didactic purposes, into three steps: conflict graph generation, cost analysis, and label selection. In actual implementation these three tasks need not be performed separately. Rather, using the trellis strategy, the cost of each conflicting label candidate can be accumulated, the derived expense for the target candidates determined, and the optimal label candidate selected, all with one swift pass through the neighboring features.

## 5. EXPERIMENTAL RESULTS

The algorithm was applied to several data views generated by the Starlight Visualization System [34]. These views had a dimension of 770x840 pixels. Table 3 and 4 represents the time required to compute a full-map labeling for various (uniform) label sizes. The times recorded are the results of the author's implementation of the algorithm (in C++) on a Dell Xeon, 3.2Ghz machine running Windows, with 3GB of RAM. See Figure 17 for an example of a typically dense point cloud, along with its labels. The use of progressively smaller label sizes demonstrates the ability of the algorithm to calculate all map resolutions or zoom levels. Note that the choice of label size does not significantly increase the time required for computation. This was true regardless of the ratio of label-size to map. Table 4 displays the results for small labels, some smaller than a pixel. The purpose of this will be discussed further below, but it essentially demonstrates the ability of the system to process all zoom levels of an enormous map—without time penalty.

---

[2] It should be noted that using both of the above shortcuts in tandem can have one somewhat unexpected drawback: It is possible (although rare in the author's implementation) for a selected label of $f_B$ to occlude the feature $f_A$ altogether. This will only occur if all of $f_A$'s candidates had been previously occluded. As a result, a label of a low-ranking feature may overpost a high-ranking feature (although only until the user has zoomed or panned sufficiently to "release" one of the higher-ranking feature's candidates). This undesirable behavior may motivate the implementer to modify or remove one of these shortcuts. On the other hand, however, this anomaly occurred with less than 0.01% of the features in the author's own test cases, and therefore is probably not a significant concern.

Table 3: **Labeling Speed (in seconds)**

| | | Label dimensions (w x h, in pixels) | | | |
|---|---|---|---|---|---|
| | | *50x8* | *100x10* | *150x12* | *200x14* |
| **Number of features** | **1K** | < 0.001 | < 0.001 | < 0.001 | < 0.001 |
| | **3K** | 0.031 | 0.031 | 0.031 | 0.032 |
| | **5K** | 0.047 | 0.047 | 0.047 | 0.047 |
| | **11K** | 0.11 | 0.11 | 0.109 | 0.109 |
| | **25K** | 0.266 | 0.328 | 0.328 | 0.344 |
| | **50K** | 0.531 | 0.625 | 0.641 | 0.641 |
| | **75K** | 0.844 | 0.984 | 1.047 | 1.063 |

These results compare favorably against all previously published methodologies. In terms of speed, this algorithm appears to be orders of magnitude faster than most other approaches, including those documented in [12, 26, 36, 43, 47]. Furthermore, these results include tests ranging up to 130,000 features, whereas no published results to date include tests with datasets larger than 20,000 features, with most less than 5000. (The implementation in [2] mentions a dataset of 12 million, but a pre-processing phase is required and no timed results are offered).

Table 4: **Labeling Speed for Much Smaller Labels**

| | Label dimensions (w x h, in pixels) | | |
|---|---|---|---|
| | *16x4* | *3x1* | *1.0x0.4* |
| **11K pts** | 0.11 | 0.11 | 0.093 |

In order to compare this algorithm with previously reported approaches, it was also applied to some of the benchmark data available at [44]. It is important to note, however, that direct comparisons are difficult due to a fundamental design difference, *viz.*: nearly all previous approaches assume a fixed-size map resolution and allow no features to be obscured. In contrast, this paper has described an algorithm suitable for maps of varying-resolutions, in which point-density *necessitates* feature overposting on most zoom levels. Moreover, previous approaches typically measured the effectiveness of an algorithm by the number of features labeled in the final solution. This metric is clearly unsuitable, however, when applied to a map that is so dense that the vast majority of its points cannot possibly be labeled. Nevertheless, some comparisons are indeed useful. The fastest recorded times previously reported in the literature appear in [36]. Although the authors of that approach do not specify the speed of the system it was run on, these results, tabulated in Table 5, indicate that the trellis strategy may be up to ten times faster. Note that this is true even though the trellis algorithm was processing data sets that were sometimes ten times larger.

**Table 5: Benchmark Data**

| | US Cities | | |
|---|---|---|---|
| | *No. of sites* | *# of labels* | *time (in sec)* |
| Algorithm in [36] | 1,041 | 1,041 | 0.5 |
| Trellis algorithm | 10,296 | 1,537 | 0.078 |
| Trellis algorithm | 100,000 | 6,883 | 0.86 |
| Trellis algorithm | 130,000 | 63,269 | 1.3 |

| | Munich Drill Holes | | |
|---|---|---|---|
| | *No. of sites* | *# of labels* | *time (in sec)* |
| Algorithm in [36] | 19,461 | 11,049 | 7.1452 |
| Trellis algorithm | 19,446 | 11,142 | 0.204 |

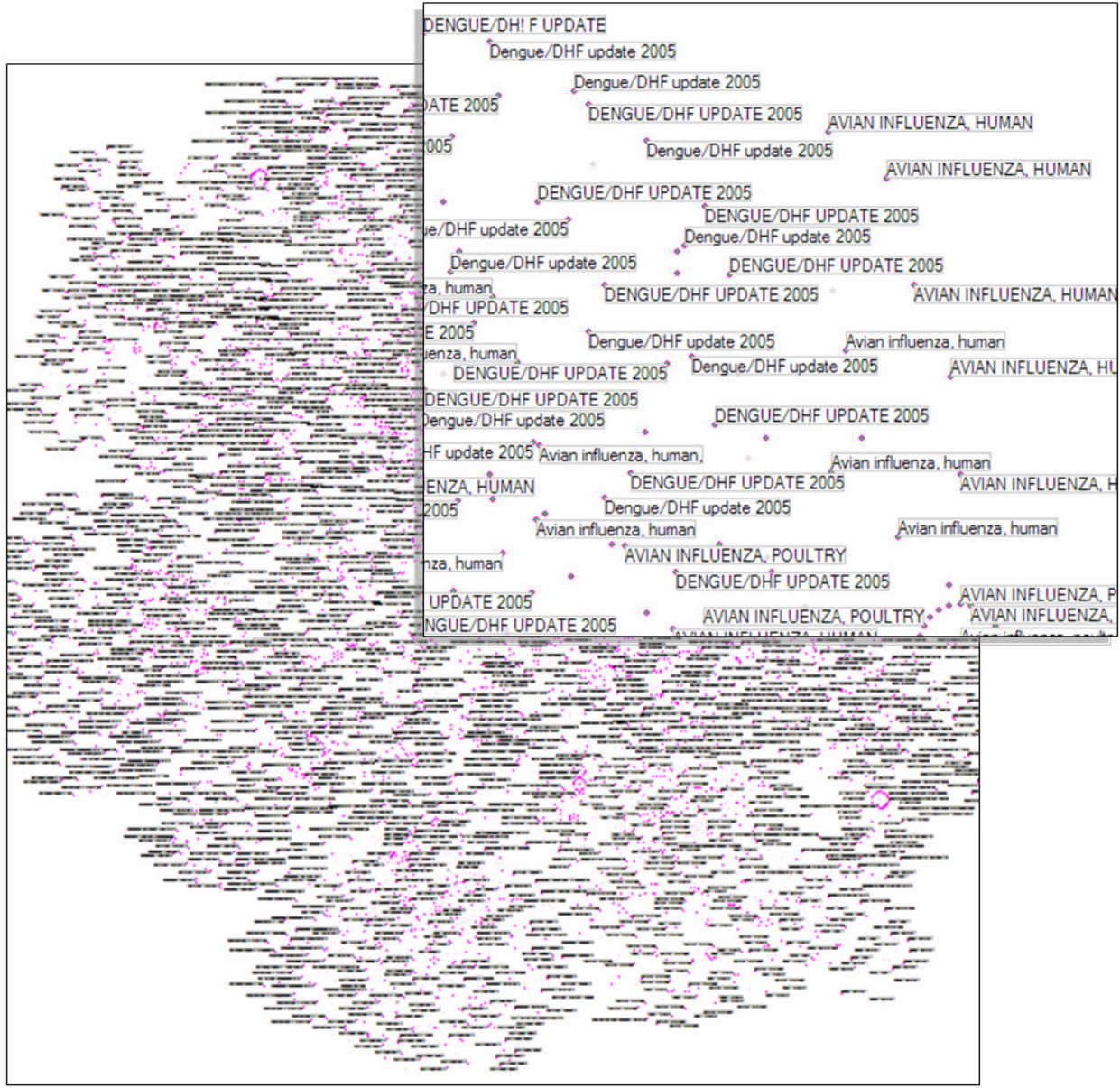

**Figure 17: 11,000 feature points from a cluster graph.** At the given resolution 3,605 labels were placed in 0.11 seconds. This particular view is not meant for display, but demonstrates the calculated label positions of one "zoom level." Eight other such zoom levels can be calculated in under a second, allowing the user to zoom in until all labels are deconflicted. The inset shows what the user would see at the current zoom level.

## 6. CONCLUSION AND SUGGESTIONS FOR FURTHER STUDY

This paper has presented a new approach for automated feature label de-confliction operating at speeds and scalability that are well-suited for interactive visual analytic applications. An algorithm was offered that begins to realize Dr. Freeman's dream of a "1-second quality labeling of an electronically displayed map" [16]. Indeed, the approach outlined here provides real-time, whole-map labeling at speeds measured in milliseconds, without the need for preprocessing. Moreover, this method has demonstrated an ability to scale, in sub-second time, to massive data sets, larger than what most previous approaches have even attempted to handle. This is a critical feature at a time when visual analytic applications routinely process tens of thousands of nodes in a single view.

The speed and scalability of this algorithm opens the door to a number of options in terms of 2D and 3D interactivity. For the two-dimensional case, two distinct modes of operation are possible:

*"Just-in-time" view de-confliction.* The original motivation for this study was to produce a de-confliction algorithm fast enough to operate at a rate of multiple frames per second. This goal has indeed been realized with data sets in excess of 25,000 nodes. At these speeds a view can be efficiently labeled and re-labeled at every interactive movement of a user's mouse. This was, in fact, the original design of the algorithm: for any given orientation of the data, the *current view* can be labeled and any previous labeling can be discarded. One of the drawbacks of this approach, however, is its failure to satisfy the so-called "desiderata," or rules of label consistency (outlined in [2]), which seek to eliminate unexpected "popping" and re-sorting of label locations. For this reason, a second option can be considered:

*Multi-level pre-processing.* This paper has, from the outset, presented pre-processing as inappropriate for the ad hoc nature of visual analytic maps. Nevertheless, the speed of this algorithm allows us to reconsider that opinion. Recall that camera zooming can be considered equivalent to a universal expansion of intra-point distances along with label-size scaling. Therefore, by pre-scaling the labels to progressively smaller sizes, the layout configuration may be computed for every zoom level at construction time. As seen in Table 5, such computations add only minimal expense. Hence, in 3-5 seconds, an entire map could be pre-processed, and its label locations stored, for up to 8 or 12 zoom levels. In so doing, the undesirable "popping" of labels that would otherwise occur as the user pans around in lower zoom levels would be eliminated. The added memory demands can be addressed through subdivision of the trellis, and the small amount of extra time required would in many cases be dwarfed by the demands of view construction in general. This would also free up cycles that may be better spent by the expensive rendering engine during view interaction. It should be said that, while this would prevent any "popping" during horizontal movement (panning), other measures are necessary to mitigate popping between zoom levels. One option in this regard would be to "lock" label locations in place, once they have been determined at a higher zoom level. Such a decision would expose the unavoidable trade-off between label consistency and label-count maximization.

Beyond the advantages offered for labeling in the two-dimensional case, this algorithm may also be applicable in the more demanding arena of three-dimensional views. Heretofore, no labeling algorithm has offered the speed required to handle the complexities of interactive 3D label de-confliction. When a user interacts with a three-dimensional view, the relative orientation and configuration of the features in the view are constantly in flux: the 2D zooming and panning capabilities are now supplemented with rotation (both camera and view), view angle manipulation, and dollying. The possible number of view orientations is literally countless and, hence, a pre-processing phase is unfeasible. Yet, by projecting the features to the view-plane, the multi-frame-per-second rate of this algorithm may finally provide real-time labels for three-dimensional views.

Another area for future study presents itself: Because the trellis strategy operates only on specific regions within a two-dimensional array, it readily lends itself to a threaded, parallel-processing approach. In fact, the algorithm may well be classified as an "embarrassingly parallel" problem—easily separable into independent tasks. See [17], particularly Chapter 8, for a description of how this might be accomplished. But even without that, it is clear, that for a vast number of applications within the arena of information visualization, the trellis strategy of label de-confliction is a fast, reliable, and worthwhile tool.

## APPENDIX A – IMPLEMENTATION DETAILS

My algorithm was built in Visual C++ within an MFC environment. This provided a simple graphical interface to interact with the program and to visually inspect the results of the algorithm. The interface itself consists of a toolbar by which a variety of parameters may be set, a main view window displaying the points and the resulting labels, and a modified histogram window, providing a rough estimate of the efficiency of the algorithm.

## A.1  USER INTERFACE

### A.1.1  PARAMETERS (TOOLBAR)

The parameters of the algorithm, as represented by the buttons on the prototype toolbar, went through an evolutionary process. The original intent of the application was to compare the relative (averaged) speeds of various published algorithms. For that reason, there are several toolbar controls which allow the user to select various algorithms, and to set their related parameters. In the testing/evaluation phase of this project, however, it was concluded that all the algorithms were woefully inadequate and that a paradigm shift in strategies was necessary. The resulting algorthmic re-engineering led to the Trellis algorithm as described in this paper. The application has a variety of controls that reflect both approaches. The various controls will now be explained. Refer to Figure 18 for the remainder of this section.

#### A.1.1.1  ORIGINAL DESIGN

In the center of the toolbar, a group of five (originally four) radio buttons allow the user to select one of four primary algorithms to be tested: Greedy, Greedy/FIFO, Three Rules, and a hybrid of Three Rules with Greedy. At the outset of this project, it was apparent from the literature that the Three Rules algorithm held the greatest promise of success. The datasets used with these tests were randomly generated dot clusters. The controls at the top of the toolbar allowed the user to select the size and density of these clusters. Two check-boxes also allow the user to display the feature number associated with each dot in the view window, and also the size of the dots themselves could be reduced (from circles, the default, to single pixel dots). One other parameter, represented by the "Options" radio button group (near the bottom), allowed the user to decide between "fast" and "thorough." This latter set of options controlled the amount of recursion desired. One last set of controls at the bottom of the toolbar allowed the user to run the algorithm multiple times on the same dataset and average the times, to get a better sense of the real time requirements of each approach.

### A.1.1.2   MODIFIED DESIGN

Once it became apparent that the original strategies were insufficient, it was decided that the Trellis Algorithm would replace the previous attempts. This prompted several modifications to the user interface. First of all, it was clear that randomly generated points were inadequate for comparison with published timing records, so the corresponding controls were abandoned. Several new controls were added which correspond to the parameters that are unique to the Trellis Algorithm, *viz.*, "Threshold" (representing $\theta$, the priority threshold), "Cell Width" (which sets the trellis cell width to be equal to the largest label size, or the average label size), and "Allowed overlap %" (which scales the trellis cell width incrementally, and in so doing specifies the amount of potential overlap to be allowed between neighboring labels). Finally, a check-box provides the option of displaying the trellis itself, and a "Toggle Label" button allows the user to turn on and off individual labels according to their feature number.

### A.1.2 OUTPUT WINDOWS

In order to interact with the data and visually inspect the results of the algorithm, two view windows were created. Refer to the following figure as necessary.

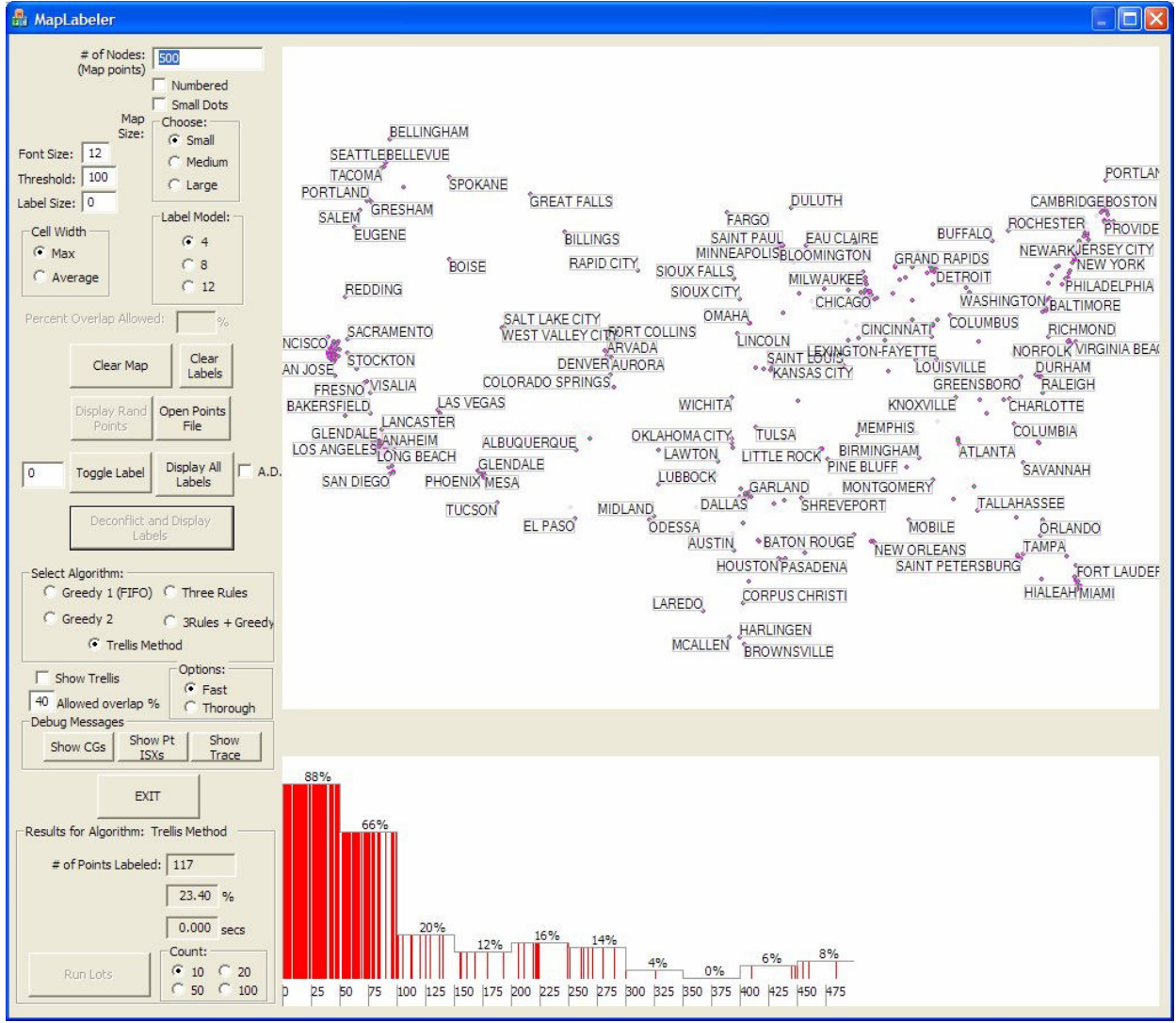

**Figure 18: The MapLabeler application interface.**

### A.1.2.1 MAIN VIEW AREA

The primary output window is capable of displaying all the features (i.e., points) in the data set. These points can be displayed as a small circle or as a one-pixel dot. The trellis can be displayed (optionally), along with a numeric index for convenience. The labels can be displayed in various ways: (1) The entire set of label candidates can be displayed. In most sets this will be a mass of indistinguishable rectangles. (2) The label candidates for a single specified point can be displayed. It will look like a grid of four rectangles centered on the specified point. (3) The final set of selected labels can be displayed, including the associated text. In this final case, the labels will look like text strings, bounded by a faint rectangle. Any obscured points will be displayed dimly (to give the effect of partially transparent labels. These labels, of course, will be attached at one of their corners to the referent point. Panning and zooming are not yet implemented in this prototype.

### A.1.2.2    HISTOGRAM

In order to offer a means by which to grade the relative effectiveness of the labeling process, a modified histogram appears immediately below the main view. Recall that most previous algorithms have been graded by the final number of labels attached to points. As discussed in this paper, this metric is not useful for the specific problem of dense point clouds in interactive displays. For a point set of 10,000 points, for example, it is clearly impossible to label even a significant fraction of the points in standard screen space. Therefore, the percent of labeled points will almost always be exceedingly low. Nevertheless, some metric needs to be offered. It is particularly important to confirm that the highest priority points are receiving the most labels (as specified by this paper), with the lower ranking points receiving a diminishing number. For this purpose the histogram divides the point set into ten slices (ordered by priority), and visually displays the "score" for each percentile.

There are two dimensions of information in the histogram. Each pixel in the horizontal dimension represents the presence of a label for a given point (with one pixel per map point). If there are more map points than there are pixels in the width of the window, the pixels are "subdivided" by "stacking" the map point data. Hence, if a given pixel represented three map points, two of which were labeled, the corresponding pixel tower would reach two-thirds of the way to its ceiling. The vertical dimension of the histogram represents the ceiling of the given percentile (the ten percent "slice" of the data). If 90% of the labels of the ten most important points in a set of 100 were selected, the ceiling of that percentile would reach 9/10 of the way to the top of the window.

Thus, in a typical labeling, the histogram will display a stair-stepped assortment of vertical lines. In a desirable labeling run, the stairs should start high and descend as necessary.

## A.2  CLASS STRUCTURE

### A.2.1        CLASS DIAGRAM

The following class diagram (Figure 19) provides a high level view of the most important aspects of the program. As it is displayed in this diagram, the MapLabelerDlg class acts as the primary interface between the user interface and the labeling engine. It handles all user interactions with the toolbar and controls. It owns a single instance of a MapSet class. The MapSet is the primary owner of all the primary data structures, including the array of all the points (MapPoints), all the labels (MapLabels), and the Trellis object. The Trellis class contains all the primary algorithmic utilities used in the label deconfliction routine. Details about the specific implementation of the application follows.

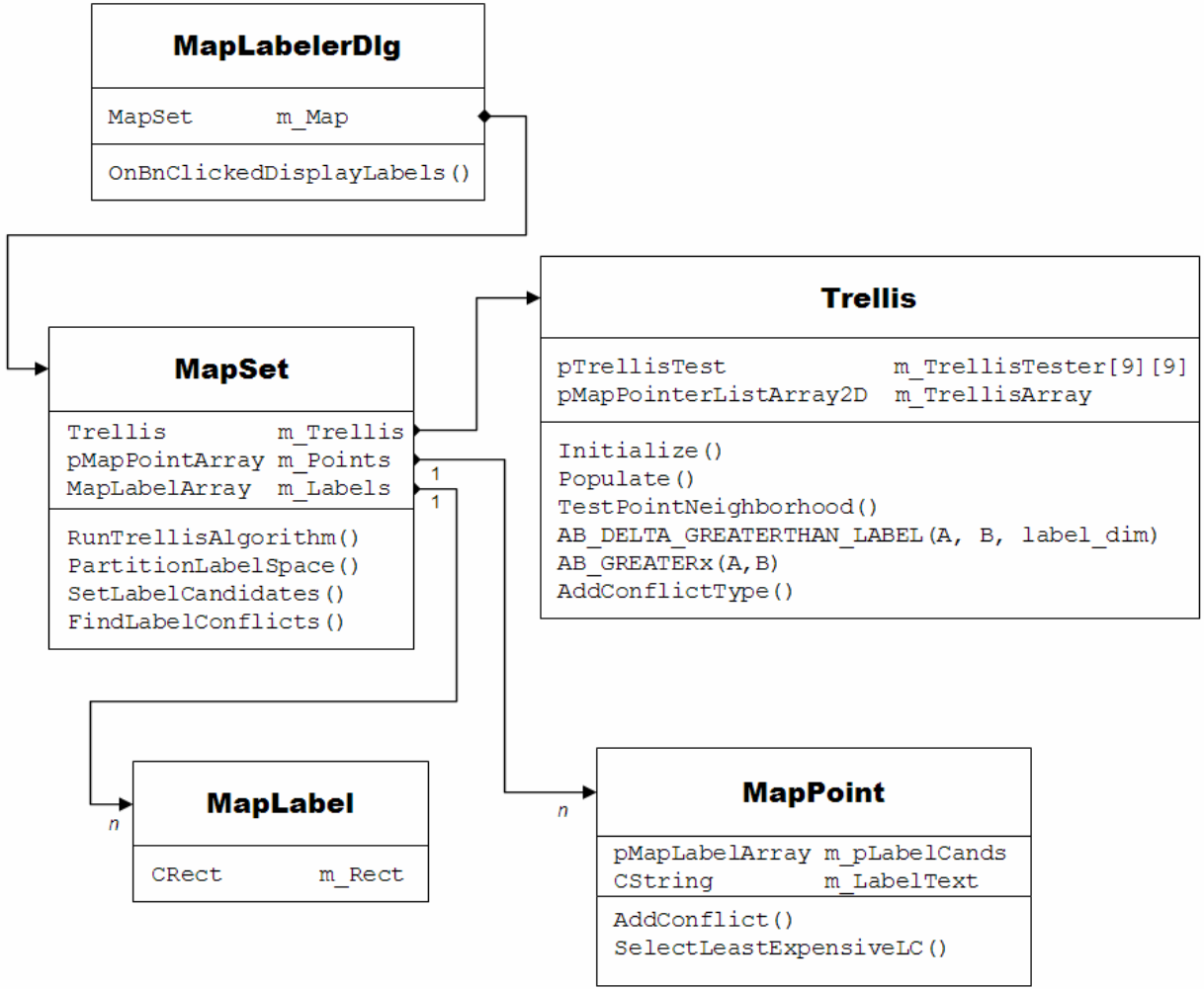

**Figure 19: The MapLabeler class diagram.**

As an interface class, the MapLabelerDlg class simply acts as a go-between to connect the user interface (toolbar) with the labeling engine. As such, it provides all the control handlers for the various toolbar controls, as well as the drawing handler for the main views (although the main functionality of the latter is provided through function calls to the MapSet object). The primary function, of course, is the handler for the button labeled "Deconflict and Display Labels." This handler simply determines the algorithm desired and calls two functions in the MapSet object: `SetLabelCandidates()` and `FindLabelConflicts(userAlgorithm)`, sending the desired algorithm as a parameter to the latter function. Once that function returns, the work is done, so the main canvas and the histogram pane are triggered for a repaint.

### A.2.3    MAPSET CLASS

As stated previously, the MapSet class is the single owner of all the primary data structures, including the array of all the points (MapPoints), all the labels (MapLabels), and the Trellis object. It is also responsible for the actual rendering of the points and labels in the main view. It is driven by the MapLabelerDlg class. The first call from that class is to the `SetUpMapFromFile()` function (in response to the "Open Points File") button. This opens a file dialog allowing the user to browse to a valid points file. A valid file is an xml document in the following format:

```
<?xml version="1.0"?>
<ViewData>
 <Data>
  <Feature_Points nodes="1000" width="522" height="380">
     <point rank="1" key1="NEW YORK" key2="NEW YORK" data="7333253"
        lat="40.71416" lon="-74.006386" x="0.8587972" y="0.2525217"/>
     <point rank="2" key1="LOS ANGELES" key2="CALIFORNIA" data="3448613"
        lat="34.05222" lon="-118.242775" x="0.6383598" y="0.2526796"/>
     <point rank="998" key1="WATERTOWN" key2="MASSACHUSETTS" data="31437"
        lat="42.37083" lon="-71.183334" x="0.6622894" y="0.2992986"/>
     <point rank="999" key1="CARMEL" key2="INDIANA" data="31411" lat="39.97833"
        lon="-86.118057" x="0.6872756" y="0.299356"/>
     <point rank="1000" key1="MANKATO" key2="MINNESOTA" data="31404"
        lat="44.16361" lon="-93.999161" x="0.6899804" y="0.2993776"/>
     .
     .
     .
  </Feature_Points>
 </Data>
</ViewData>
```

Once a valid file has been opened and parsed, the MapSet populates its two feature arrays: the `pMapPointArray` and the `MapLabelArray`. (The latter array is stored separately from the former simply for convenience during the drawing function.

Once the MapSet is populated it is ready for operation. In response to a "Deconflict and Display Labels" button event the MapLabelerDlg class calls the `SetLabelCandidates()` function. This function simply loops through all the points and calls `PartitionLabelSpace()`, passing the appropriate label model (which is always 4 in the current implementation).

`PartitionLabelSpace()` instantiates four label candidates for the given point, establishing their height and width and orientation relative to the referent point.

Once the MapSet has been thus initialized, the MapLabelerDlg is free to call the primary function `FindLabelConflicts()` which simply acts as a switch statement selecting the appropriate algorithm to run. For the purposes of this paper, the function called is `RunTrellisAlgorithm()`.

Central to this class is the `RunTrellisAlgorithm()` function, which is divided into three primary stages. Stage One initializes the trellis with the canvas rectangle (i.e., the screen size), and the individual "quarter-region" cell dimensions. It then populates the trellis data structure by sending a pointer to the MapPoint array. Stage Two loops through all the points in the MapPoint array (or the first $\theta$ of them, if a threshold is applied) and calls the `TestPointNeighborhood()` function on the Trellis object. Stages One and Two are bounded by a system timer. The timer is set prior to the beginning of Stage One and it is terminated when the loop in Stage Two is complete. The final stage in this function serves to calculate the data necessary to build and draw the histogram, providing visual metrics for the user to inspect.

### A.2.4 TRELLIS CLASS

The Trellis Class has a number of unique features which were added both for efficiency at runtime as well as clarity and ease of maintenance. The two primary edifices are the main trellis data structure, and the trellis tester pointer array.

The primary data structure `m_TrellisArray`, is a two-dimensional array of MapPoint lists. The lists are built on the `CTypedPtrList` provided by MFC. When the `Trellis::Populate()` function is called by the MapSet (passing a MapPoint array), the Trellis loops through each MapPoint in the array and calculates which quarter-region the point resides in using the following calculations:

```
int map_point_col = (int)(floor(a_Point.x/(float)m_qrtRgnWidth));
int map_point_row = (int)(floor(a_Point.y/(float)m_qrtRgnHeight));³
```

Once the trellis coordinates have been determined, the points are added to a list in the given array cell.

The other distinguishing feature of the Trellis class is a unique device called the `m_TrellisTester`. This structure is a (9x9) array of function pointers. It operates as a sliding "screen" of function calls which completely tests the points in the neighborhood of a given referent point. The notion here is based upon the structure of the main "Trellis Neighborhood Test Outline" (cf. Table 1 and Appendix B). This outline carefully delineates the exact tests that are prescribed for each "block" in the 9x9 neighborhood of a point. The particular test prescribed depends on the geographic relationship to the home cell (e.g., "three cells up and two cells to the left"). All points in the neighbor cell are subject to the same hierarchy of tests. Therefore,

---

³ It is interesting to note that although this computation is valid for horizontally aligned labels, the user can choose instead to display the labels at a specified angle, and the only change necessary as far as the deconfliction engine is concerned is an alteration to this region calculation. To display the labels at an angle φ, this simple substitution is required:

For `a_Point.x` we substitute `Point.x/cos(φ) + dx`,
and for `a_Point.y` we substitute `a_Point.y/cos((π/2) – φ) + dy`.

where `dx` and `dy` can be calculated trigonometrically based on the dimensions of the screen space. These calculations are outside the scope of this paper, however, as this suggested modification has not been implemented yet.

`m_TrellisTester` is used to easily loop through the neighborhood and apply the appropriate test functions.

By way of example, consider the function pointed to by `m_TrellisTester[0][0]` (the upper left cell in the neighborhood). This function is called `Test_n4_n4()` because it represents the cell with the coordinates (-4, -4) in the Test Outline. This function receives a list of points from the corresponding trellis array, loops through each one and performs the following computations (where `pointA` is the referent point, and `pointB` is a given point in neighbor cell (-4, -4)):

```
{
   if (pointA->m_Rank > pointB->m_Rank) continue;
      //don't test higher-ranked features
   if ((!AB_DELTAy2(pointA, pointB, m_AvgLabelWidth, m_AvgLabelHeight ))
      &&(!AB_DELTAx2(pointA, pointB, m_AvgLabelWidth, m_AvgLabelHeight )))
         AddConflictType(pointA, pointB, beta1, a_NeigborhoodRadius);
}
```

and where `AB_DELTAx2` and `AB_DELTAy2` are macros which perform the following tests, respectively:

```
Abs(fA.x_coord - fB.x_coord) > 2*Label_width
Abs(fA.y_coord - fB.y_coord) > 2*Label_height
```

which returns true if `pointB` is more than two label widths away from `pointA` in the *x*-direction (respectively for *y*). The function `AddConflictType()` is an extended switch statement, with enumerated case types `beta0` through `delta3`, which correspond to the particular conflict configurations delineated in Table 2. Depending on the particular configuration, one or more conflicts are added to the referent point with a call to `MapPoint::AddConflictT()`. This MapPoint function is also responsible for determining the cost of a given collision, as explained in Section A.2.5 below.

The primary driving function within the Trellis class is `TestPointNeighborhood()`, called by the MapSet. As its name suggests, this function is responsible for applying the Trellis Tester to the neighborhood of a single passed MapPoint. The only additional complexity that must be considered is that if a MapPoint resides near the border of the global trellis (that is, if the `NEIGHBOR_RADIUS` is greater than the distance to the edge of the trellis border), not all of its neighboring cells might exist. In this case, only a portion of the Trellis Tester must be applied.

Once the neighborhood has been tested in entirety, a final label candidate is selected by a call to `SelectLeastExpensiveLC` (), described in the next section.

### A.2.5        MAPPOINT CLASS

The MapPoint class has only two primary tasks.

The function `AddConflictT(pointA, pointB)`, records the existence of a collision between two points. In addition, it determines the "expense" of the given conflict as a sum of the conflict partner's inherent cost multiplied by a "zoom factor." The zoom factor is an additional weighting determined by the radial distance from the central cell. The purpose of this, as described in Section 4.2.2, is to allow the cost to be adjusted upwardly for conflicts that occur in more nearly located cell neighbors. Finally, this weighted conflict cost is added to the total expense of the given candidates of `pointA`.

The second task of this class is embodied in `SelectLeastExpensiveLC()`. This selecting function simply loops through all four label candidates and determines which of the non-occluded candidates have the lowest associated cost. If two candidates are equally expensive, it returns the highest index (as per the label-model preference criteria).

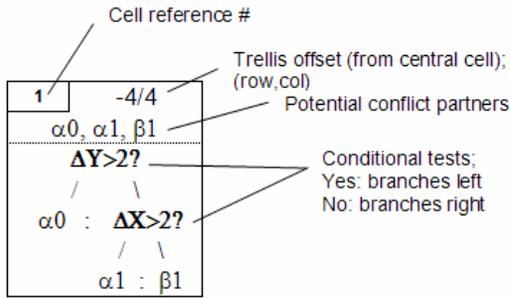

Cell reference #

Trellis offset (from central cell); (row,col)

Potential conflict partners

Conditional tests;
Yes: branches left
No: branches right

| 1 | -4/4 |
| α0, α1, β1 | |
| ΔY>2? | |

α0 : ΔX>2?

α1 : β1

## APPENDIX B – TRELLIS NEIGHBORHOOD TEST OUTLINE

**Conditional Tests:**

| | |
|---|---|
| **ΔY>1?** | → abs($f_A$.y_coord - $f_B$.y_coord) > Label_height |
| **ΔY>2?** | → abs($f_A$.y_coord - $f_B$.y_coord) > 2*Label_height |
| **$Y_A$>$Y_B$?** | → $f_A$.y_coord > $f_B$.y_coord |
| | |
| **ΔX>1?** | → abs($f_A$.x_coord - $f_B$.x_coord) > Label_width |
| **ΔX>2?** | → abs($f_A$.x_coord - $f_B$.x_coord) > 2*Label_width |
| **$X_A$>$X_B$?** | → $f_A$.x_coord > $f_B$.x_coord |

| 1 -4/-4 | 2 -4/-3 | 3 -4/-2 | 4 -4/-1 | 5 -4/0 | 6 -4/1 | 7 -4/2 | 8 -4/3 | 9 -4/4 |
|---|---|---|---|---|---|---|---|---|
| α0, α1, β1 | α1, β1 | α1, β1, γ13 | α1, γ13 | α1, γ13, γ31 | α1, γ31 | α1, β3, γ31 | α1, β3 | α1, α3, β3 |
| ΔY>2? | | ΔY>2? | | ΔY>2? | | ΔY>2? | | ΔY>2? |
| / \ | ΔY>2? | / \ | ΔY>2? | / \ | ΔY>2? | / \ | ΔY>2? | / \ |
| α0 : ΔX>2? | / \ | α1 : ΔX>1? | / \ | α1 : $X_A$>$X_B$? | / \ | α1 : ΔX>1? | / \ | α1 : ΔX>2? |
| / \ | α1 : β1 | / \ | α1 : γ13 | / \ | α1 : γ31 | / \ | α1 : β3 | / \ |
| α1 : β1 | | β1 : γ13 | | γ13 : γ31 | | β3 : γ31 | | α3 : β3 |
| 10 -3/-4 | 11 -3/-3 | 12 -3/-2 | 13 -3/-1 | 14 -3/0 | 15 -3/1 | 16 -3/2 | 17 -3/3 | 18 -3/4 |
| α0, β1 | β1 | β1, γ13 | γ13 | γ13, γ31 | γ31 | β3, γ31 | β3 | α3, β3 |
| ΔX>2? | | ΔX>1? | | $X_A$>$X_B$? | | ΔX>1? | | ΔX>2? |
| / \ | No test | / \ | No test | / \ | No test | / \ | No test | / \ |
| α0 : β1 | | β1 : γ13 | | γ13 : γ31 | | β3 : γ31 | | α3 : β3 |
| 19 -2/-4 | 20 -2/-3 | 21 -2/-2 | 22 -2/-1 | 23 -2/0 | 24 -2/1 | 25 -2/2 | 26 -2/3 | 27 -2/4 |
| α0, β1, γ10 | β1, γ10 | β1, γ10, γ13, δ1 | γ13, δ1 | γ13, γ31, δ1, δ3 | γ31, δ3 | β3, γ31, γ32, δ3 | β3, γ32 | α3, β3, γ32 |
| ΔX>2? | | ΔX>1? | | ΔY>1? | | ΔX>1? | | ΔX>2? |
| / \ | ΔY>1? | / \ | ΔY>1? | / \ | ΔY>1? | / \ | ΔY>1? | / \ |
| α0 : ΔY>1? | / \ | ΔY>1? : ΔY>1? | / \ | $X_A$>$X_B$? : $X_A$>$X_B$? | / \ | ΔY>1? : ΔY>1? | / \ | α3 : ΔY>1? |
| / \ | β1 : γ10 | / \ / \ | γ13 : δ1 | / \ / \ | γ31 : δ3 | / \ / \ | β3 : γ32 | / \ |
| β1 : γ10 | | β1 : γ10 γ13 : δ1 | | γ13 : γ31 δ1 : δ3 | | β3 : γ32 γ31 : δ3 | | β3 : γ32 |
| 28 -1/-4 | 29 -1/-3 | 30 -1/-2 | 31 -1/-1 | 32 -1/0 | 33 -1/1 | 34 -1/2 | 35 -1/3 | 36 -1/4 |
| α0, γ10 | γ10 | γ10, δ1 | δ1 | δ1, δ3 | δ3 | γ32, δ3 | γ32 | α3, γ32 |
| ΔX>2? | | ΔX>1? | | $X_A$>$X_B$? | | ΔX>1? | | ΔX>2? |
| / \ | No test | / \ | No test | / \ | No test | / \ | No test | / \ |
| α0 : γ10 | | γ10 : δ1 | | δ1 : δ3 | | γ32 : δ3 | | α3 : γ32 |
| 37 0/-4 | 38 0/-3 | 39 0/-2 | 40 0/-1 | 41 0/0 | 42 0/1 | 43 0/2 | 44 0/3 | 45 0/4 |
| α0, γ10, γ01 | γ01, γ10 | γ01, γ10, δ0, δ1 | δ0, δ1 | δ0, δ1, δ2, δ3 | δ2, δ3 | γ23, γ32, δ2, δ3 | γ23, γ32 | α3, γ23, γ32 |
| ΔX>2? | | ΔX>1? | | $X_A$>$X_B$? | | ΔX>1? | | ΔX>2? |
| / \ | $Y_A$>$Y_B$? | / \ | $Y_A$>$Y_B$? | / \ | $Y_A$>$Y_B$? | / \ | $Y_A$>$Y_B$? | / \ |
| α0 : $Y_A$>$Y_B$? | / \ | $Y_A$>$Y_B$? : $Y_A$>$Y_B$? | / \ | $Y_A$>$Y_B$? : $Y_A$>$Y_B$? | / \ | $Y_A$>$Y_B$? : $Y_A$>$Y_B$? | / \ | α3 : $Y_A$>$Y_B$? |
| / \ | γ01 : γ10 | / \ / \ | δ0 : δ1 | / \ / \ | δ2 : δ3 | / \ / \ | γ23 : γ32 | / \ |
| γ01 : γ10 | | γ01 : γ10 δ0 : δ1 | | δ0 : δ1 δ2 : δ3 | | γ23 : γ32 δ2 : δ3 | | γ23 : γ32 |

| 46 — 1/-4 | 47 — 1/-3 | 48 — 1/-2 | 49 — 1/-1 | 50 — 1/0 | 51 — 1/1 | 52 — 1/2 | 53 — 1/3 | 54 — 1/4 |
|---|---|---|---|---|---|---|---|---|
| $\alpha0, \gamma01$ | $\gamma01$ | $\gamma01, \delta0$ | $\delta0$ | $\delta0, \delta2$ | $\delta2$ | $\gamma23, \delta2$ | $\gamma23$ | $\alpha3, \gamma23$ |
| $\Delta X>2?$ / \ $\alpha0 : \gamma01$ | No test | $\Delta X>1?$ / \ $\gamma01 : \delta0$ | No test | $X_A>X_B?$ / \ $\delta0 : \delta2$ | No test | $\Delta X>1?$ / \ $\gamma23 : \delta2$ | No test | $\Delta X>2?$ / \ $\alpha3 : \gamma23$ |

| 55 — 2/-4 | 56 — 2/-3 | 57 — 2/-2 | 58 — 2/-1 | 59 — 2/0 | 60 — 2/1 | 61 — 2/2 | 62 — 2/3 | 63 — 2/4 |
|---|---|---|---|---|---|---|---|---|
| $\alpha0, \beta0, \gamma01$ | $\beta0, \gamma01$ | $\beta0, \gamma01, \gamma02, \delta0$ | $\gamma02, \delta0$ | $\gamma02, \gamma20, \delta0, \delta2$ | $\gamma20, \delta2$ | $\beta2, \gamma20, \gamma23, \delta2$ | $\beta2, \gamma23$ | $\alpha3, \beta2, \gamma23$ |
| $\Delta X>2?$ / \ $\alpha0 : \Delta Y>1?$ / \ $\beta0 : \gamma01$ | $\Delta Y>1?$ / \ $\beta0 : \gamma01$ | $\Delta X>1?$ / \ $\Delta Y>1? : \Delta Y>1?$ / \ / \ $\beta0 : \gamma01$  $\gamma02 : \delta0$ | $\Delta Y>1?$ / \ $\gamma02 : \delta0$ | $\Delta Y>1?$ / \ $X_A>X_B? : X_A>X_B?$ / \ / \ $\gamma02 : \gamma20$  $\delta0 : \delta2$ | $\Delta Y>1?$ / \ $\gamma20 : \delta2$ | $\Delta X>1?$ / \ $\Delta Y>1? : \Delta Y>1?$ / \ / \ $\beta2 : \gamma23$  $\gamma20 : \delta2$ | $\Delta Y>1?$ / \ $\beta2 : \gamma23$ | $\Delta X>2?$ / \ $\alpha0 : \Delta Y>1?$ / \ $\beta2 : \gamma23$ |

| 64 — 3/-4 | 65 — 3/-3 | 66 — 3/-2 | 67 — 3/-1 | 68 — 3/0 | 69 — 3/1 | 70 — 3/2 | 71 — 3/3 | 72 — 3/4 |
|---|---|---|---|---|---|---|---|---|
| $\alpha0, \beta0$ | $\beta0$ | $\beta0, \gamma02$ | $\gamma02$ | $\gamma02, \gamma20$ | $\gamma20$ | $\beta2, \gamma20$ | $\beta2$ | $\alpha3, \beta2$ |
| $\Delta X>2?$ / \ $\alpha0 : \beta0$ | No test | $\Delta X>1?$ / \ $\beta0 : \gamma02$ | No test | $X_A>X_B?$ / \ $\gamma02 : \gamma20$ | No test | $\Delta X>1?$ / \ $\beta2 : \gamma20$ | No test | $\Delta X>2?$ / \ $\alpha3 : \beta2$ |

| 73 — 4/-4 | 74 — 4/-3 | 75 — 4/-2 | 76 — 4/-1 | 77 — 4/0 | 78 — 4/1 | 79 — 4/2 | 80 — 4/3 | 81 — 4/4 |
|---|---|---|---|---|---|---|---|---|
| $\alpha2, \alpha0, \beta0$ | $\alpha2, \beta0$ | $\alpha2, \beta0, \gamma02$ | $\alpha2, \gamma02$ | $\alpha2, \gamma02, \gamma20$ | $\alpha2, \gamma20$ | $\alpha2, \beta2, \gamma20$ | $\alpha2, \beta2$ | $\alpha2, \alpha3, \beta2$ |
| $\Delta Y>2?$ / \ $\alpha2 : \Delta X>2?$ / \ $\alpha0 : \beta0$ | $\Delta Y>2?$ / \ $\alpha2 : \beta0$ | $\Delta Y>2?$ / \ $\alpha2 : \Delta X>1?$ / \ $\beta0 : \gamma02$ | $\Delta Y>2?$ / \ $\alpha2 : \gamma02$ | $\Delta Y>2?$ / \ $\alpha2 : X_A>X_B?$ / \ $\gamma02 : \gamma20$ | $\Delta Y>2?$ / \ $\alpha2 : \gamma20$ | $\Delta Y>2?$ / \ $\alpha2 : \Delta X>1?$ / \ $\beta2 : \gamma20$ | $\Delta Y>2?$ / \ $\alpha2 : \beta2$ | $\Delta Y>2?$ / \ $\alpha2 : \Delta X>2?$ / \ $\alpha3 : \beta2$ |

## REFERENCES

[1]  Agarwal, P. K., van Kreveld, M., and Suri, S., "Label placement by maximum independent set in rectangles," *Computational Geometry: Theory and Applications*, vol. 11, no. 3-4, pp. 209-218, 1998.

[2]  Been, K., Daiches, E., and Yap, C., "Dynamic Map Labeling," *IEEE Transactions on Visualization and Computer Graphics*, vol. 12, no. 5, pp. 773-780, 2006.

[3]  Bekos, M. A., Kaufmann, M., and Tae-Cheon, Y., "Labeling collinear sites," in *Proc.Asia-Pacific Symp.Information Visualization* IEEE, 2007, pp. 45-51.

[4]  Chazelle, B. and Amenta, N., "Application Challenges to Computational Geometry," Computational Geometry Impact Task Force,TR-521-96, 1999.

[5]  Christensen, J., Marks, J., and Shieber, S., "Labeling point features on maps and diagrams," *Technical Report TR-25-92*, 1992.

[6]  Christensen, J., Marks, J., and Shieber, S., "An empirical study of algorithms for point-feature label placement," *ACM Transactions on Graphics*, vol. 14, no. 3, p. 203, 1995.

[7]  Christophe, G. and Philippe, R., "The Sort/Sweep Algorithm: A New Method for R-tree Based Spatial Joins," in *Proceedings of the 12th International Conference on Scientific and Statistical Database Management (SSDBM'00)* IEEE Computer Society, 2000, pp. 153-165.

[8]  Cravo, G. M., Ribeiro, G. M., and Lorena, L. A. N., "A greedy randomized adaptive search procedure for the point-feature cartographic label placement," *Computers & Geosciences*, p. -(To appear), 2007.

[9]  Doddi, S., Marathe, M. V., Mirzaian, A., Moret, B. M. E., and Zhu, B., "Map labeling and its generalizations," in *Proceedings of the Annual ACM-SIAM Symposium on Discrete Algorithms* 1997, p. 148.

[10] Doddi, S., Marathe, M. V., and Moret, B. M. E., "Point set labeling with specified positions," in *Proceedings of the Annual Symposium on Computational Geometry* 2000, p. 182.

[11] Dorschlag, D., Petzold, I., and Plümer, L., "Placing Objects Automatically in Areas of Maps," in *Proc.23rd International Cartographic Conference (ICC'03)* 2003.

[12] Ebner, D., Klau, G., and Weiskircher, R., "Label number maximization in the slider model," in *Proc. 12th Internat. Symp. on Graph Drawing (GD'04)*. Pach, J., Ed. New York: Institut für Computergraphik und Algorithmen, Technische Univ. Wien, 2004, pp. 144-154.

[13] Edmondson, S., Christensen, J., Marks, J., and Shieber, S., "A general cartographic labeling algorithm," *Cartographica*, vol. 33, no. 4, pp. 13-23, 1997.

[14] Fekete, J. and Plaisant, C., "Excentric labeling: Dynamic neighborhood labeling for data visualization," in *Proc.Conference on Human Factors in Computer Systems (CHI'99)*, Pittsburgh PA: 1999, pp. 512-519.

[15] Formann, M. and Wagner, F., "A packing problem with applications to lettering of maps," in *Proceedings of the seventh annual symposium on Computational geometry,* North Conway, New Hampshire, United States: ACM Press, 1991, pp. 281-288.

[16] Freeman, H., "Automated cartographic text placement," *Pattern Recognition Letters*, vol. 26, no. 3, pp. 287-298, 2005.

[17] Grama, A., Gupta, A Karypis G, and Kumar, V, *Introduction to Parallel Computing*. Harlow, England: Addison-Wesley, 2003.

[18] Guting, R. H. and Schilling, W., "A practical divide-and-conquer algorithm for the rectangle intersection problem," *Information Sciences*, vol. 42, no. 2, pp. 95-112, 1987.

[19] Harrie, L., Stigmar, H., Koivula, T., and Lehto, L., "An Algorithm for Icon Labelling on a Real-Time Map," in *Proc.11th International Symposium on Spatial Data Handling,* Heidelberg: Springer-Verlag, 2005, pp. 493-507.

[20] Hartmann, K., Götzelmann, T., Ali, K., and Strothotte, T., "Metrics for Functional and Aesthetic Label Layouts," in *Smart Graphics, 5th International Symposium* 2005 pp. 115-126.

[21] Imhof, E., "Positioning names on maps," *The American Cartographer*, vol. 2, no. 2, pp. 128-144, 1975.

[22] Kakoulis, K. G. and Tollis, I. G., "On the edge label placement problem," in *Proceedings of the Symposium on Graph Drawing,* Nice, Fr: ACM, New York, NY, USA, 1997, pp. 453-465.

[23] Kakoulis, K. G. and Tollis, I. G., "A unified approach to automatic label placement," *Int. J. Comput. Geom. Appl.*, vol. 13, no. 1, pp. 23-60, 2003.

[24] Kakoulis, K. G. and Tollis, I. G., "Unified approach to labeling graphical features," in *Proceedings of the Annual Symposium on Computational Geometry* 1998, pp. 347-384.

[25] Kirkpatrick, S., Gelatt, C. D. J., and M.P.Vecchi, "Optimization by simulated annealing," *Science*, vol. 220, no. 4598, pp. 671-680, 1983.

[26] Lorena, L. and Ribeiro, G., "Heuristics for cartographic label placement problems.," *Computers and GeoSciences*, vol. 32, no. 6, pp. 739-748, 2006.

[27] Marks, J. and Shieber, S., "The Computational Complexity of Cartographic Label Placement," Harvard CS,Technical Report 05-91, Mar.1991.

[28] Oosterom, P., "Reactive Data-Structures for Geographic Information Systems." PhD Thesis, Department for Computer Science, Leiden University, Netherlands, 1990.

[29] Petzold, I., Gröger, G., and Plümer, L., "Modeling of Conflicts for Screen Map Labeling," in *Proceedings of the XXth ISPRS Congress, Istanbul, Turkey 2004,* Istanbul, Turkey: 2004.

[30] Petzold, I., Plümer, L., and Gröger, G., "Fast screen map labeling - data-structures and algorithms," in *Proc.23rd International Cartographic Conference (ICC'03)* 2003.

[31] Petzold, I., Plümer, L., and Heber, M., "Label placement for dynamically generated screen maps," in *Proc.19th International Cartographic Conference (ICC'99)* 1999, pp. 893-903.

[32] Poon, S. H., Shin, C. S., Strijk, T., Uno, T., and Wolff, A., "Labeling points with weights," *Algorithmica (New York)*, vol. 38, no. 2, pp. 341-362, 2003.

[33] Poon, S. H. and Shin, C. S., "Adaptive Zooming in Point Set Labeling," in *Proc.15th International Symposium on Fundamentals of Computation Theory (FCT),* Lübeck, Germany: 3623 Springer 2005, 2005, pp. 222-233.

[34] Risch, J., Rex, D., Dowson, S., Walters, T., May, R., and Moon, B., "The STARLIGHT information visualization system," in *Readings in information visualization: using vision to think*. Card, S., Mackinlay, J., and Shneiderman, B., Eds. San Francisco: Morgan Kaufmann Publishers Inc, 1999, pp. 551-560.

[35] Rostamabadi, F. and Ghodsi, M., "Unit height k-position map labeling," in *Proc.19th European Workshop on Computational Geometry (EWCG'03),* Bonn: 2003.

[36] Roy, S., Bhattacharjee, S., Das, S., and Nandy, S. C., "A fast algorithm for point labeling problem," in *Proc.17th Canad.Conf.Computational Geometry (CCCG)* 2005, pp. 155-158.

[37] Skupin, A., "The world of geography: Visualizing a knowledge domain with cartographic means," *Proceedings of the National Academy of Sciences*, vol. 101, no. suppl_1, pp. 5274-5278, Apr.2004.

[38] Strijk, T. and van Kreveld, M., "Practical extensions of point labeling in the slider model," in *ACM Workshop on Advances in Geographic Information Systems,* Hong Kong, Hong Kong: Association for Computing Machinery, New York, NY, USA, 1999, p. 190.

[39] Strijk, T., Verweij, B., and Aardal, K., "Algorithms for maximum independent set applied to map labelling," Dept of Comp Sci, Utrecht Univ.,Technical Report UU-CS-2000-22, 2000.

[40] Thomas, J. and Cook, K eds, *Illuminating the Path -- The Research and Development Agenda for Visual Analytics,* IEEE Computer Society Press, 2006.

[41] van Kreveld, M., Strijk, T., and Wolff, A., "Point labeling with sliding labels," *Comput. Geom. -Theory Appl.*, vol. 13, no. 1, pp. 21-47, 1999.

[42] Wagner, F. and Wolff, A., "A combinatorial framework for map labeling," in *Graph Drawing.6th International Symposium, GD'98.Proceedings,* Montreal, Que., Canada: Springer-Verlag, 1998, p. 316.

[43] Wagner, F., Wolff, A., Kapoor, V., and Strijk, T., "Three Rules Suffice for Good Label Placement," *Algorithmica (New York)*, vol. 30, no. 2, pp. 334-346, 2001.

[44] Wolff, A. and Strijk, T., "Map Labeling Bibliography," http://i11www.iti.uni-karlsruhe.de/map-labeling/bibliography, updated: Feb.2007.

[45] Wong, P. C., Mackey, P., Perrine, K., Eagan, J., Foote, H., and Thomas, J., "Dynamic Visualization of Graphs with Extended Labels," in *Proceedings of the 2005 IEEE Symposium on Information Visualization* IEEE Computer Society, 2005, pp. 73-80.

[46] Yamamoto, M., Camara, G., and Lorena, L., "Fast Point-Feature Label Placement Algorithm for Real Time Screen Maps," in *VI Brazilian Symposium in Geoinformatics* 2005.

[47] Zhang, Q. and Harrie, L., "Placing text and icon labels simultaneously: A real-time method.," *Cartography and Geographic Information Science*, vol. 33, no. 1, pp. 53-64, 2006.

[48] Zhang, Q. and Harrie, L., "Real-Time Map Labeling for Personal Navigation," in *Proc.12th Int.Conf.on Geoinformatics - Geospatial Information Research: Bridging the Pacific and Atlantic,* University of Gävle, Sweden: 2004.

[49] Zhu, B. and Qin, Z. P., "New Approximation Algorithms for Map Labeling with Sliding Labels," *Journal of Combinatorial Optimization*, vol. 6, no. 1, pp. 99-110, 2002.